\documentstyle[12pt]{article}
\def\ie{{\em i.e.}}
\def\eg{{\em e.g.}}

\def\beq{\begin{equation}}
\def\eeq{\end{equation}}

\catcode`\@=11 % This allows us to modify PLAIN macros.
\def\coeff#1#2{{\textstyle{#1\over #2}}}

\def\VEV#1{\left\langle #1\right\rangle}
\def\vev#1{\langle #1\rangle}
\def\lsim{\mathrel{\mathpalette\@versim<}}
\def\gsim{\mathrel{\mathpalette\@versim>}}
\def\@versim#1#2{\vcenter{\offinterlineskip
    \ialign{$\m@th#1\hfil##\hfil$\crcr#2\crcr\sim\crcr } }}
\def\etal{{\em et. al.}}
\def\JL{J. L. Lopez}
\def\DVN{D. V. Nanopoulos}
\def\AZ{A. Zichichi}

\def\r#1{$\bf#1$}
\def\rb#1{$\bf\overline{#1}$}

\def\t1{{\tilde 1}}

\def\F{\widetilde F}
\def\Fb{\widetilde{\bar F}}

\def\eV{\,{\rm eV}}

\def\GeV{\,{\rm GeV}}

\def\y{\,{\rm y}}

\def\to{\rightarrow}
\def\pb{\,{\rm pb}}
\def\ipb{\,{\rm pb}^{-1}}

\def\h{\coeff{1}{2}}
\def\rt{\coeff{1}{\sqrt{2}}}
\def\NPB#1#2#3{Nucl. Phys. B {\bf#1} (19#2) #3}
\def\PLB#1#2#3{Phys. Lett. B {\bf#1} (19#2) #3}

\def\PRD#1#2#3{Phys. Rev. D {\bf#1} (19#2) #3}
\def\PRL#1#2#3{Phys. Rev. Lett. {\bf#1} (19#2) #3}

\def\MODA#1#2#3{Mod. Phys. Lett. A {\bf#1} (19#2) #3}
\def\IJMP#1#2#3{Int. J. Mod. Phys. A {\bf#1} (19#2) #3}
\def\TAMU#1{CTP-TAMU-#1}

\def\hepph#1{({\tt hep-ph/#1})}
\def\hepth#1{({\tt hep-th/#1})}

\begin{document}
\begin{flushright}
\baselineskip=12pt
CTP-TAMU-41/95\\
DOE/ER/40717--19\\
ACT-15/95\\
\tt hep-ph/9511266
\end{flushright}

\begin{center}
\vglue 1cm
{\Large\bf Flipped SU(5): a Grand Unified Superstring Theory (GUST)
Prototype\\}
\vglue 0.75cm
{\Large Jorge L. Lopez$^1$ and D.V. Nanopoulos$^{2,3}$\\}
\vglue 0.5cm
\begin{flushleft}
$^1$Department of Physics, Bonner Nuclear Lab, Rice University\\ 6100 Main
Street, Houston, TX 77005, USA\\
$^2$Center for Theoretical Physics, Department of Physics, Texas A\&M
University\\ College Station, TX 77843--4242, USA\\
$^3$Astroparticle Physics Group, Houston Advanced Research Center (HARC)\\
The Mitchell Campus, The Woodlands, TX 77381, USA\\
\end{flushleft}
\end{center}

\vglue 0.75cm
\begin{abstract}
In this Lecture we first review the basic properties that make flipped SU(5) a
very economical and well motivated unified model, including some recent
developments regarding the prediction for $\alpha_s$ and new implications for
proton decay. Then we sketch the derivation of flipped SU(5) from strings,
stressing some new results concerning the cosmological constant, the stability
of the no-scale mechanism, and a new mechanism for generating the ``LEP" scale
$M_{\rm LEP}\sim10^{16}\GeV$ in string models. Finally we present a sample
supersymmetry breaking scenario, where all sparticle masses depend on a single
parameter, and discuss the experimental signatures at LEP and the Tevatron.
This scenario also predicts the top-quark mass to be close to 175 GeV.
\end{abstract}
\vspace{1cm}
\begin{flushleft}
\baselineskip=12pt
CTP-TAMU-41/95\\
DOE/ER/40717--19\\
ACT-15/95\\
November 1995
\end{flushleft}
\footnotetext{Lecture presented by D.~V.~Nanopoulos at the 33rd International
School of Subnuclear Physics ``Vacuum and vacua: the physics of nothing",
Erice, July 2-10, 1995.}

\newpage
\tableofcontents
\newpage

\setcounter{page}{1}
\pagestyle{plain}
\baselineskip=14pt

\section{Standard Flipped SU(5)}
\label{sec:standard}
\subsection{Status of GUTs}
\label{subsec:status}
\begin{figure}[t]
\vspace{3in}
\includegraphics{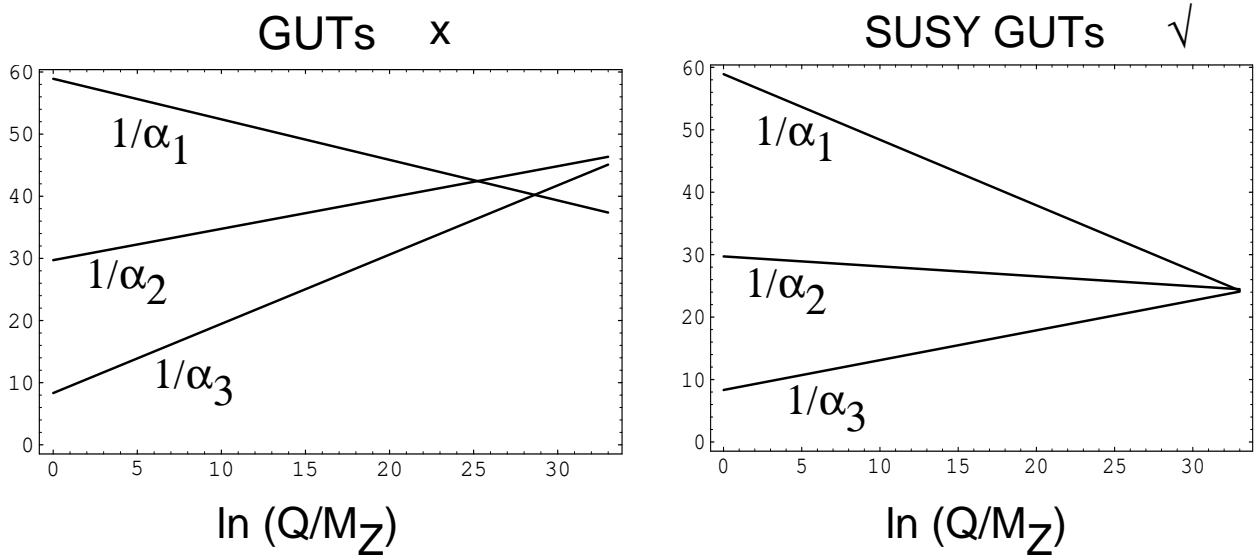}
\caption{Running of the Standard Model gauge couplings in minimal GUTs and
SUSY GUTs, showing the failure of the former and the success of the latter.}
\label{fig:runnings}
\end{figure}

In an nutshell, the status of GUTs can be summarized as follows. LEP has
measured the Standard Model gauge couplings with unprecedented precision,
leading to the following world averages
\begin{eqnarray}
\sin^2\theta_W&=&0.23143\pm0.00028\ , \\
\alpha_s(M_Z)&=&0.118\pm0.006\ .
\end{eqnarray}
These couplings run with energy scale as prescribed by the renormalization
group equations (RGEs), and may or may not converge at a single point. In the
simplest GUTs without supersymmetry (\eg, the Georgi-Glashow SU(5) model
\cite{GG}) the gauge couplings do not converge \cite{amaldi-costa} and thus
fail the defining property of GUTs; a result that confirmed their earlier
demise due to their incorrect proton lifetime prediction. Moreover, simple or
complicated GUTs without supersymmetry are undesirable, as they fail to resolve
the gauge hierarchy problem, or break dynamically the electroweak symmetry via
radiative corrections. On the other hand, in the simplest SUSY GUTs the gauge
couplings do converge around $M_{\rm GUT}\sim10^{16}\GeV$ \cite{EKN}. This
contrast is illustrated in Fig.~\ref{fig:runnings}. The question then becomes
\begin{quotation}
Are all GUTs with supersymmetry equally viable?
\end{quotation}
The answer to this question has to be no, as can be illustrated in the case
of the minimal SU(5) SUSY GUT, which has several shortcomings. Some of these
tend to be overlooked, as the problem with the doublet-triplet splitting, the
lack of neutrino masses, and the lack of a mechanism for baryogenesis. More
recently it has become apparent that its prediction for $\alpha_s(M_Z)$
\cite{baggeretal}
\begin{equation}
{\rm SU(5):}\qquad\alpha_s(M_Z)>0.123
\end{equation}
is very problematic. These troubles are tabulated below, and contrasted with
their counterparts in Flipped SU(5) \cite{Barr,revitalized}.
\begin{table}[h]
\caption{Comparison between SU(5) and flipped SU(5) GUT features.}
\label{Table1}
\begin{center}
\begin{tabular}{|l|c|c|}\hline
Basic GUT tests&SU(5)&Flipped SU(5)\\ \hline
$\sin^2\theta_W\Rightarrow\alpha_3(M_Z)$&$\times$&$\surd$\\
Proton decay&$p\to \bar\nu K^+$&$p\to e^+\pi^0$\\
Doublet-triplet splitting&$\times$&$\surd$\\
Neutrino masses&$\times$&$\surd$\\
Baryogenesis&$\times$&$\surd$\\ \hline
\end{tabular}
\end{center}
\vspace{-0.6cm}
\end{table}
Some people like to perform ``model independent" analyses by ignoring the GUT
structure altogether, as is the case in ``MSSM" studies. This makes little
sense, as the seed of the low-energy model predictions is found at or above
the GUT scale. For instance, the constraint from proton decay is many times
neglected, as if it could be satisfied automatically. Yet, in GUT models where
this does happen (as in flipped SU(5) \cite{revitalized}) the GUT structure is
quite different from that assumed in these ``generic" models.
When considering different SUSY GUT models, one can also ask the question
\begin{quotation}
What \underline{really} is the ``minimal" GUT?
\end{quotation}
The analysis of Georgi and Glashow indicated that the smallest GUT structure
containing the non-abelian Standard Model gauge groups is SU(5). Judging by
the number of group generators, one would then consider flipped SU(5), and
afterwards SO(10). However, we would argue that SU(5) is out of the running, as
it lacks the desirable features mentioned above. Flipped SU(5) is the next
larger gauge structure, and thus is the ``minimal" SUSY GUT incorporating all
the desirable GUT features mentioned above.\footnote{These features can be
incorporated in the missing-doublet version (MDM) of SU(5) \cite{MPM}, but this
model is quite cumbersome and hardly ``minimal" \cite{MDM}.}  Flipped SU(5)
differs from SU(5)
and SO(10) in a very important respect: the electric charge generator does
not lie completely within SU(5), as the gauge group is really
SU(5)$\times$U(1). We would like to argue that such ``non-grand" unification of
the non-abelian Standard Model gauge couplings is in fact desirable, as the
complete ``grand" unification of all interactions must also include the
gravitational sector (\ie, the hidden sector and the supersymmetry breaking
sector), and this is predicted not to occur until the string scale. To make
the relevance of this argument more apparent, we have contrasted the cases
of unified and grand unified SU(2) and SU(5) gauge groups in
Table~\ref{Table2}.

\begin{table}[t]
\caption{Comparison between unified and grand unified SU(2) and SU(5) gauge
groups and their properties.}
\label{Table2}
\begin{center}
\vspace{-0.7cm}
\begin{tabular}{|l|l|}\hline
\underline{SU(2)}&\underline{SU(2)$\times$U(1)}\\
{\small[Georgi-Glashow `72]}&{\small[Glashow `61, Weinberg `67, Salam `68]}\\
$\bullet$ ``grand" unified&$\bullet$ unified\\
$\bullet$ $W^\pm,\gamma$: $\gamma$ inside SU(2)&$\bullet$ $W^\pm,Z,\gamma$:
$\gamma=\{W^3$ [SU(2)], $B$ [U(1)]$\}$\\
$\bullet$ Higgs triplet (adjoint)&$\bullet$ Higgs doublet (antisymmetric)\\
$\bullet$ Neutral currents exist (1973)&$\bullet$ SU(3) not accounted for;
grand unification later\\
$\bullet$ Wrong!&$\bullet$ Right!\\ \hline\hline
\underline{SU(5)}&\underline{SU(5)$\times$U(1)}\\
{\small[Georgi-Glashow `74]}&{\small[Barr `82, Derendinger-Kim-Nanopoulos
`84,}\\
&{\small Antoniadis-Ellis-Hagelin-Nanopoulos `87]}\\
$\bullet$ grand unified&$\bullet$ unified\\
$\bullet$ $W^\pm,W^3,B,X,Y$&$\bullet$ $W^\pm,W^3,B,X,Y,\widetilde B$\\
$\quad \gamma$ inside SU(5)&$\quad\gamma$: $(W^3,B)$ [SU(5)], $\widetilde B$
[U(1)]\\
$\bullet$ Higgs \r{24} (adjoint)&$\bullet$ Higgs \r{10},\rb{10}
(antisymmetric)\\
$\bullet$ $\alpha_3>0.123$ $M_{\rm GUT}\sim {1\over100} M_{\rm Pl}$&$\bullet$
Gravity$^*$ not accounted for; grand unification later\\
$\bullet$ Wrong?&$\bullet$ Right?\\ \hline
\multicolumn{2}{l}{\small$^*$ ``Gravity" = supersymmetry breaking, hidden
sector gauge groups, string unification.}
\end{tabular}
\end{center}
\end{table}

\subsection{Basic flipped SU(5) features}
\label{subsec:basic}
Because of the matter field content and gauge structure in flipped SU(5),
several desirable GUT mechanisms are present \cite{revitalized} and work rather
naturally, as we know briefly review.
\subsubsection{Matter fields} The model includes three generations of quarks
and leptons; (\r{10},\rb{10}) Higgs GUT representations whose
$\nu^c_H,\nu^c_{\bar H}$ components break the SU(5)$\times$U(1) GUT symmetry
down to the Standard Model once they acquire suitable vevs; and a pair of Higgs
pentaplets which include the two light Higgs doublets. Some singlet fields
($\phi$) are also present.
\begin{description}
\item $F_{(10)}=\{Q,d^c,\nu^c\}$; $\bar f_{(\bar 5)}=\{L,u^c\}$; $l_{(1)}=e^c$
\ (3 generations)
\item $H_{(10)}=\{Q_H,d^c_H,\nu^c_H\}$; $\bar H_{(\overline{10})}=\{Q_{\bar
H},d^c_{\bar H},\nu^c_{\bar H}\}$
\item $h_{(5)}=\{H_2,H_3\}$; $\bar h_{(\bar5)}=\{\bar H_2,\bar H_3\}$
\end{description}
\subsubsection{GUT superpotential}
As allowed by the SU(5)$\times$U(1) gauge symmetry, GUT couplings include
interactions to effect the missing-partner mechanism and the see-saw mechanism
\begin{equation}
W_G=\lambda_4 H H h+\lambda_5 \bar H\bar H\bar h+\lambda_6 F\bar H\phi
+\mu h\bar h
\end{equation}
\subsubsection{Doublet-triplet splitting}
The components of the Higgs pentaplets must be split
\begin{equation}
h=\left(
\begin{array}{c}
H_2\\
H_3
\end{array}
\right)
\begin{array}{l}
{\rm electroweak\ symmetry\ breaking}\\
{\rm proton\ decay}
\end{array}
\end{equation}
because of their very different roles. The interactions in $W_G$
\begin{eqnarray}
&\lambda_4 HHh\to \lambda_4d^c_H\vev{\nu^c_H}H_3\\
&\lambda_5\bar H\bar H\bar h\to\lambda_5\bar d^c_H\vev{\bar\nu^c_H}\bar H_3
\end{eqnarray}
make the triplets heavy, while leaving the doublets light (the missing
partner mechanism). A similar mechanism in the missing-doublet extension of
SU(5) (MDM) requires the introduction of large representations
(\r{50},\rb{50},\r{75}) for this sole purpose \cite{MPM}.
\subsubsection{Yukawa superpotential}
The ``flipping" of the assignments of the Standard Model fields to the SU(5)
representations entails the couplings
\begin{equation}
\lambda_u F\bar f\,\bar h+\lambda_d FFh+\lambda_e\bar fl^ch\ .
\end{equation}
Note that the usual $\lambda_b=\lambda_\tau$ relation in SU(5) is replaced by
$\lambda_t=\lambda_\nu$, which has consequences for the spectrum of neutrino
masses.
\subsubsection{Neutrino masses}
\label{sec:neutrinos}
 The ``flipping" mentioned above brings the $\nu^c$
field into the $F$ representation, providing a source of Dirac neutrino masses,
in addition to the see-saw type coupling in $W_G$. These interactions result
in a $3\times3$ see-saw mass matrix with the high scale provided by
$M_U=\lambda_6\vev{\nu^c_H}$,
\begin{equation}
\left.
\begin{array}{l}
\lambda_u F\bar fh\to m_u\nu\nu^c\\
\lambda_6 F\bar H\phi\to \lambda_6\vev{\nu^c_{\bar H}}\nu^c\,\phi
\end{array}
\right\}\qquad
M_\nu=
\begin{array}{c}
\nu\\ \nu^c\\ \phi
\end{array}
\stackrel{\begin{array}{ccc} \nu\quad&\nu^c&\quad\phi\end{array}}
{\left(
\begin{array}{ccc}
0&m_u&0\\ m_u&0&M_U\\ 0&M_U&M
\end{array}
\right)}
\end{equation}
The light neutrino masses then become
\begin{equation}
m_{\nu_{e,\mu,\tau}}\sim {m^2_{u,c,t}\over M^2_U/M}
\end{equation}
With the typical values $M_U\sim10^{15}\GeV$ and $M\sim10^{18}\GeV$ we
get $m_{\nu_\tau}\sim10\eV$ \cite{chorus}, which provides a very desirable
source of hot dark matter. The result for $m_{\nu_\mu}\sim10^{-3}\eV$ is
consistent with the MSW explanation for the solar neutrino deficit
\cite{chorus}. Also, the right-handed (or ``flipped") neutrino provides a
source of lepton asymmetry which is later recycled into a baryon asymmetry by
the electroweak-scale sphaleron interactions \cite{ENO}.
\subsubsection{Proton decay}
\begin{figure}[b]
\vspace{1.5in}
\caption{Dimension-five proton decay operator obtained from the $FFh$, $F\bar
f\bar h$, and $H_3\bar H_3$ interactions which, when suitably dressed by squark
and gaugino loops, mediates the decay channel $p\to\bar\nu K^+$. This channel
is negligible in flipped SU(5) because of the absence of the $H_3\bar H_3$
mixing term.}
\label{fig:d5pd}
\includegraphics{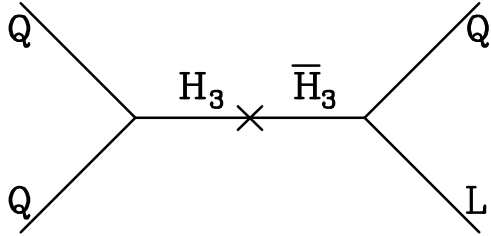}
\end{figure}

Dimension-six operators mediate proton decay via the exchange of the
$X,Y$ GUT gauge bosons, leading to the dominant decay mode $p\to e^+\pi^0$.
Whether this mode is observable at SuperKamiokande (starting in mid 1996) or
not depends on the magnitude of the unification scale. Quantitative predictions
are given in Section~\ref{subsec:alpha3} below. Proton decay via dimension-five
operators is typically the largest process in traditional GUTs, such as minimal
SU(5), where experimental limits on the mode $p\to\bar\nu K^+$ impose strict
restrictions on the allowed parameter space of the model. The elementary
diagram responsible for this reaction is shown in Fig.~\ref{fig:d5pd}.
The vertices originate from pieces of the couplings $FFh$, $F\bar f\bar h$
\begin{equation}
\lambda_d F Fh\supset QQH_3\,;\qquad
\lambda_ u F\bar f\,\bar h\supset QL\bar H_3
\end{equation}
and are therefore proportional to the Yukawa matrices. Once dressed by squark
and gaugino loops, this diagram leads to the well known $p\to\bar\nu K^+$ decay
mode. However, in flipped SU(5) this does not happen because there is no
$H_3,\bar H_3$ mixing, even though $H_3,\bar H_3$ are individually heavy via
the doublet-triplet splitting mechanism discussed above.

\subsection{Prediction for $\alpha_s(M_Z)$}
\label{subsec:alpha3}
\begin{figure}[h]
\vspace{4in}
\includegraphics{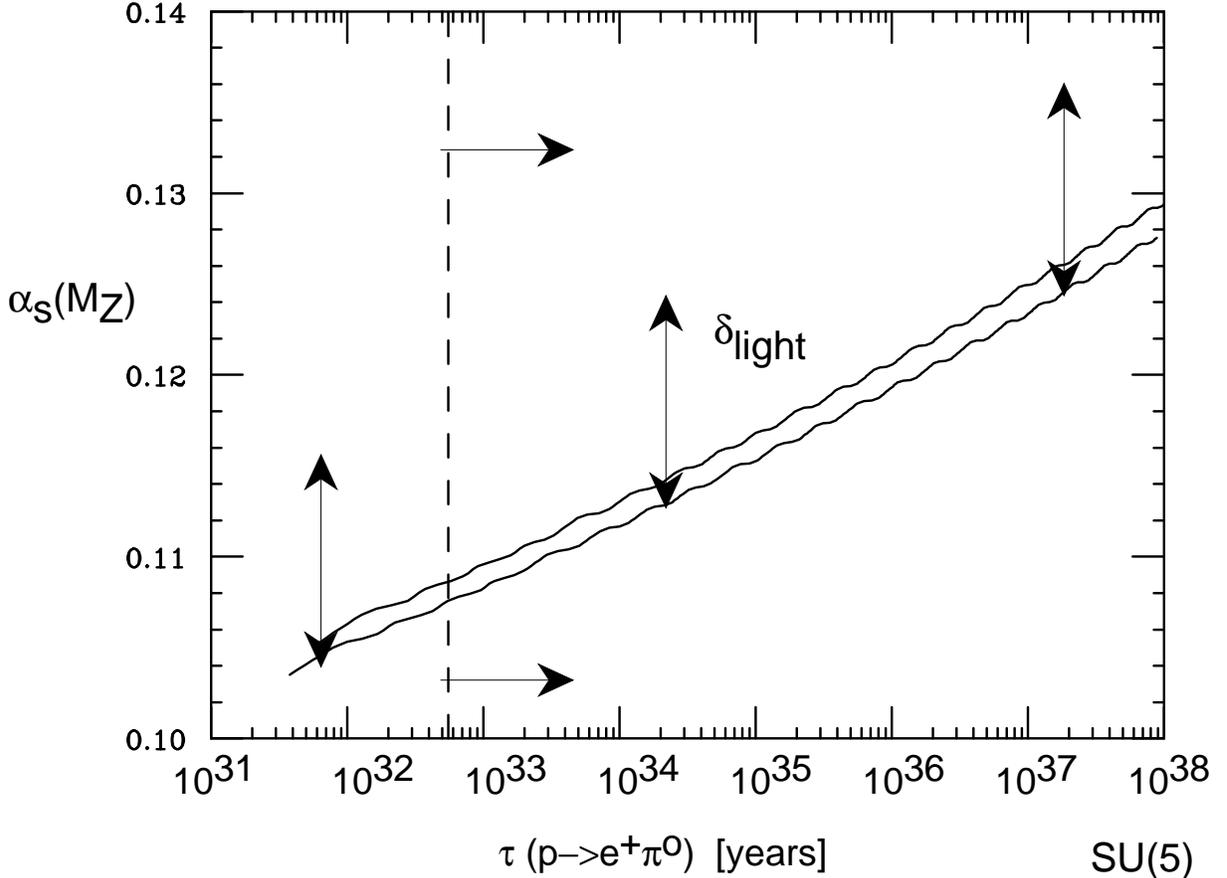}
\vspace{1.cm}
\caption{Prediction for $\alpha_s(M_Z)$ in flipped SU(5) as a function
of the proton lifetime into the mode $e^+\pi^0$. The present experimental lower
bound is indicated by the dashed line. Vertical arrows indicate the effect
of $\delta_{\rm light}$.}
\label{fig:taup}
\end{figure}

In flipped SU(5), there is a first unification scale $M_{32}$ at
which the SU(3) and SU(2) gauge couplings become equal, given to lowest order
by \cite{faspects}
\begin{eqnarray}
{1\over\alpha_2}-{1\over\alpha_5}&=&{b_2\over2\pi}\,
\ln{M_{32}\over M_Z}\ ,
\label{eq:RGE2}\\
{1\over\alpha_3}-{1\over\alpha_5}&=&{b_3\over2\pi}\,
\ln{M_{32}\over M_Z}\ ,
\label{eq:RGE3}
\end{eqnarray}
where $\alpha_2=\alpha/\sin^2\theta_W$, $\alpha_3=\alpha_s(M_Z)$, and the
one-loop beta functions are $b_2=+1$, $b_3=-3$. On the other hand, the
hypercharge gauge coupling $\alpha_Y={5\over3}
(\alpha/\cos^2\theta_W)$ evolves in general to a different
value $\alpha_1'$ at the scale $M_{32}$:
\begin{equation}
{1\over\alpha_Y}-{1\over\alpha_1'}={b_Y\over2\pi}\,
\ln{M_{32}\over M_Z}\ ,
\label{eq:RGEY}
\end{equation}
with $b_Y={33\over5}$. Above the scale $M_{32}$ the gauge group is
SU(5)$\times$U(1), with the U(1) gauge coupling $\alpha_1$ related to
$\alpha_1'$ and
the SU(5) gauge coupling ($\alpha_5$) by
\begin{equation}
{25\over\alpha_1'}={1\over\alpha_5}+{24\over\alpha_1}\ .
\label{eq:U(1)}
\end{equation}
The SU(5) and U(1) gauge couplings continue to
evolve above the scale $M_{32}$, eventually becoming
equal at a higher scale $M_{51}$. The consistency condition that
$M_{51}\ge M_{32}$ implies $\alpha_1'\le\alpha_5(M_{32})$.
The maximum possible value of $M_{32}$ is obtained
when $\alpha_1'=\alpha_5(M_{32})$ and is given by
\begin{equation}
{1\over\alpha_Y}-{1\over\alpha_5}={b_Y\over2\pi}\,
\ln{M^{\rm max}_{32}\over M_Z}\ .
\label{eq:Mumax}
\end{equation}
Solving the above equations for the value of $\alpha_s(M_Z)$ we obtain
\cite{lowering}
\begin{equation}
\alpha_s(M_Z)={\coeff{7}{3}\,\alpha\over 5\sin^2\theta_W-1
+{11\over2\pi}\,\alpha\ln(M^{\rm max}_{32}/M_{32})}\ .
\label{eq:fLO}
\end{equation}
\begin{figure}[t]
\vspace{4.0in}
\includegraphics{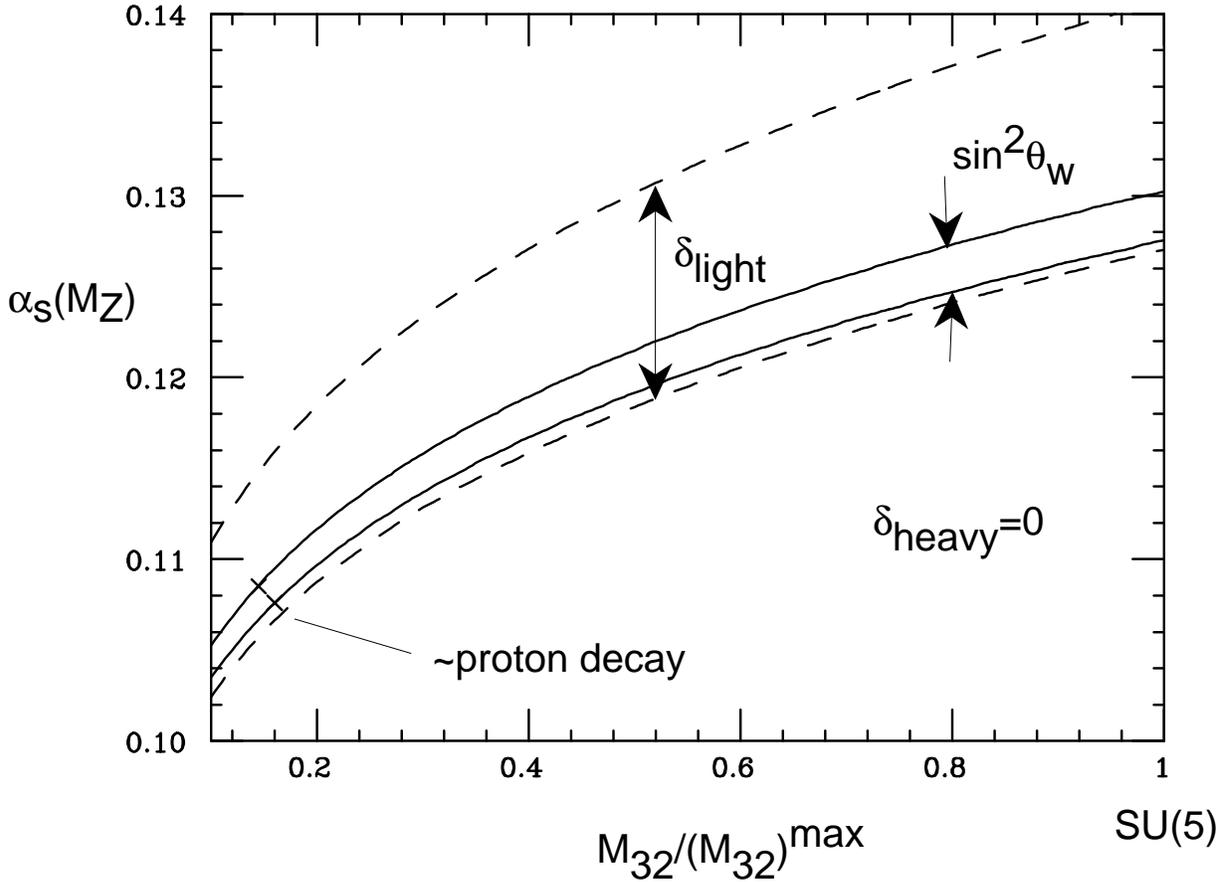}
\vspace{1.5cm}
\caption{Prediction for $\alpha_s(M_Z)$ in flipped SU(5) as a function
of $M_{32}/M^{\rm max}_{32}$. The solid lines represent the range
of predictions for $\sin^2\theta_W=0.23143\pm0.00028$ with no threshold
corrections ($\delta_{\rm light}=\delta_{\rm heavy}=0$). The dashed lines
represent the excursion obtained by scanning over sparticle masses
($\delta_{\rm light}\not=0$) below 1 TeV.}
\label{fig:sin2r}
\end{figure}
Note that since in minimal SU(5) $M_{32}=M^{\rm max}_{32}$, we automatically
obtain
\begin{equation}
\alpha_s(M_Z)^{\rm Flipped\ SU(5)}<\alpha_s(M_Z)^{\rm SU(5)}\ .
\end{equation}
The next-to-leading order corrections to Eq.~(\ref{eq:fLO}) are obtained
by the substitution
\begin{equation}
\sin^2\theta_W\to \sin^2\theta_W-\delta_{\rm 2loop}
-\delta_{\rm light}-\delta_{\rm heavy}\ ,
\label{eq:NLO}
\end{equation}
where $\delta_{\rm2loop}$ accounts for the two-loop contributions to the
RGEs, $\delta_{\rm light}$ accounts for the light SUSY thresholds, and
$\delta_{\rm heavy}$ accounts for the GUT thresholds. In both minimal SU(5)
and flipped SU(5) one finds $\delta_{\rm2loop}\approx0.0030$ and $\delta_{\rm
light}\gsim0$. These effects go the wrong way for minimal SU(5), and lead
to the lower bound $\alpha_s^{\rm SU(5)}(M_Z)>0.123$ \cite{baggeretal}. There
is no help from
$\delta_{\rm heavy}$ as it has to be positive to minimize the proton lifetime
via dimension-five operators. In the case of flipped SU(5), the reduced
unification scale allows a lower prediction for $\alpha_s(M_Z)$, and one could
even get further help from $\delta_{\rm heavy}$ \cite{lowering}
\begin{equation}
\delta_{\rm heavy}={\alpha\over20\pi}
\left[ -6\ln{M_{32}\over M_{H_3}}-6\ln{M_{32}\over M_{\bar H_3}}
+4\ln{M_{32}\over M_V}\right]\ .
\label{heavyfSU5}
\end{equation}
Since there is no problem with proton decay, the $H_3,\bar H_3$ masses can be
lighter than $M_{32}$ and $\delta_{\rm heavy}>0$ is perfectly acceptable.
In Fig.~\ref{fig:sin2r} we show the prediction for $\alpha_s(M_Z)$ in flipped
SU(5) as a function of the ratio $M_{32}/M^{\rm max}_{32}$.
We note that the minimal SU(5) prediction is obtained when $M_{32}/M^{\rm
max}_{32}=1$. Scanning over the possible range of sparticle masses
described above, we find significant variations in the predictions for
$\alpha_s(M_Z)$, indicated by the dashed lines in Fig.~\ref{fig:sin2r}, usually
towards higher values (\ie, $\delta_{\rm light}>0$).
Equations (\ref{eq:NLO}) and (\ref{eq:fLO}) show that
including the effects of
$\delta_{\rm heavy}$ simply amounts to a re-scaling
of the $M_{32}/M^{\rm max}_{32}$
axis on Fig.~\ref{fig:sin2r}, \ie,
\begin{equation}
{M_{32}\over M^{\rm max}_{32}}\to {M_{32}\over M^{\rm max}_{32}}
\ e^{-10\pi\,\delta_{\rm heavy}/11\alpha}\ .
\label{scaling}
\end{equation}
Decreasing the unification scale enhances the proton decay rate.
One obtains \cite{lowering}
\begin{equation}
\tau(p\to e^+\pi^0)\approx1.5\times10^{33}
\left({M_{32}\over10^{15}\GeV}\right)^4
\left({0.042\over\alpha_5}\right)^2 ,
\label{eq:taupf}
\end{equation}
In Fig.~\ref{fig:taup} we plot $\alpha_s(M_Z)$ versus $\tau(p\to e^+\pi^0)$.
The present experimental lower bound $\tau(p\to e^+\pi^0)^{\rm
exp}>5.5\times10^{32}\y$  allows values of $\alpha_s(M_Z)$ as low as 0.108.
Moreover, for values of $\alpha_s(M_Z)\lsim0.114$, the mode $p\to e^+\pi^0$
should be observable at SuperKamiokande in flipped SU(5), whereas in minimal
supersymmetric SU(5) the dominant mode is expected to be $p\to\bar\nu K^+$.

\section{Stringy Flipped SU(5)}
\label{sec:stringy}
The flipped SU(5) GUT model described in the previous section has many nice
features that make it a very appealing model of physics at very high energies.
However, it leaves several unanswered questions
\begin{itemize}
\item What are the particle masses (e.g., $m_t$)?
\item What are superparticle masses (e.g., $m_{\tilde q}$)?
\item What about unification with gravity?
\end{itemize}
Making supersymmetry a local symmetry (\ie, supergravity) has several
advantages. For instance, the superparticle masses can be parametrized in terms
of much fewer parameters. A popular scenario assumes universality of these
supersymmetry-breaking mass terms, resulting in a model with only four
parameters. Still, gravity is only incorporated as an effective theory below
the Planck mass. But even supergravity is not enough, as it leaves its own
set of unanswered questions
\begin{itemize}
\item What are the new supergravity functions?
\item What about gravity \underline{at} the Planck scale?
\item What about the particle masses?
\end{itemize}
These appear rather daunting questions, especially the last one
\begin{quotation}
``No theory can predict any particle mass" -- Not!
\end{quotation}
In fact, these three questions can be answered in the context of string models
although, because of the large number of possible models, the answers at the
moment are not unique. String theorists consider this to be no prediction at
all. However in practice, even though there exist many possible models, and
even some semi-realistic ones (like stringy flipped SU(5)), one can hardly
claim that the one and only string model has been found. Thus, the search for
the best string model appears to be a perfectly legitimate and physically
reasonable pursuit. (It is also possible that somehow there is not a unique
vacuum (or model) of string.)

\subsection{String derivation of Flipped SU(5)}
Flipped SU(5) was originally obtained as a string model in its so-called
``revamped" version in 1989 \cite{revamp}. The phenomenological properties of
this model were then extensively explored from first-principles string
calculations \cite{KLN,decisive}. With time it became apparent that it was
difficult to achieve unification of the gauge couplings near the string scale,
as such scenario required additional specific intermediate-scale matter
representations with Standard Model quantum numbers \cite{price}, which were
not available. In 1992 a systematic exploration of a large class of string
flipped SU(5) models \cite{search} led to a new model with a one extra pair
of (\r{10},\rb{10}) representations that in principle allowed string
unification. Here we outline the derivation and main properties of this
latest and more realistic flipped SU(5) string model \cite{search}.

As is well known, in free-fermionic string model building one must provide
a set of basis vectors of boundary conditions for the two-dimensional
world-sheet fermions as they traverse the one-loop (torus) world-sheet.
Our choice is:
\begin{eqnarray}
{\bf 1}&=(1\ 111\ 111\ 111\ 111\ 111\ 111\ :\ 111111\ 111111\ 11111\ 111
\ 1_8)\nonumber\\
S&=(1\ 100\ 100\ 100\ 100\ 100\ 100\ :\ 000000\ 000000\ 00000\ 000\ 0_8)
\nonumber\\
b_1&=(1\ 100\ 100\ 010\ 010\ 010\ 010\ :\ 001111\ 000000\ 11111\ 100\
0_8)\nonumber\\
b_2&=(1\ 010\ 010\ 100\ 100\ 001\ 001\ :\ 110000\ 000011\ 11111\ 010\
0_8)\nonumber\\
b_3&=(1\ 001\ 001\ 001\ 001\ 100\ 100\ :\ 000000\ 111100\ 11111\ 001\
0_8)\nonumber\\
b_4&=(1\ 100\ 100\ 010\ 001\ 001\ 010\ :\ 001001\ 000110\ 11111\ 100\
0_8)\nonumber\\
b_5&=(1\ 001\ 010\ 100\ 100\ 001\ 010\ :\ 010001\ 100010\ 11111\ 010\
0_8)\nonumber\\
\alpha&=(0\ 000\ 000\ 000\ 000\ 000\ 011\ :\ 000001\ 011001\ \h\h\h\h\h\
\h\h\h\ A)\nonumber
\end{eqnarray}
One also needs to specify a matrix of GSO projections which determine which
states in the spectrum are allowed by the modular invariance properties of
the two-dimensional world-sheet
\begin{equation}
k=\pmatrix{
2 &1 &2 &2 &2 &2 &2 &1\cr
1 &1 &1 &1 &1 &1 &1 &4\cr
2 &2 &2 &2 &2 &2 &2 &1\cr
2 &2 &2 &2 &2 &2 &2 &1\cr
2 &2 &2 &2 &2 &2 &2 &4\cr
2 &2 &2 &2 &2 &2 &2 &1\cr
2 &2 &2 &2 &2 &2 &2 &3\cr
2 &1 &1 &1 &1 &1 &2 &3\cr}
\end{equation}
The well-known rules of free-fermionic model-building can then be employed
to deduce a series of properties of the resulting model, such as the gauge
group, the particle spectrum, and all the interactions in the model. The gauge
group $G=G_{\rm observable}\times G_{\rm hidden}\times U(1)^5$ contains an
observable gauge group $G_{\rm observable}$=SU(5)$\times$U(1) and a hidden
gauge group $G_{\rm hidden}$=SO(10)$\times$SU(4). The particle spectrum divides
itself into a finite number of ``massless" particles and an infinite tower of
Planck mass particles. The massless fields include:

\begin{itemize}
\item Observable Sector:
\begin{equation}
\begin{array}{lll}
F^{\{0,1,2,3,4\}}[10]
&\bar f^{\{2,3,5\}}[\bar 5]&{\ell^c}^{\{2,3,5\}}[1]\\
\bar F^{\{4,5\}}[\overline{10}]\\
h^{\{1,2,3,45\}}[5]&\bar h^{\{1,2,3,45\}}[\bar5]
\end{array}
\end{equation}
\item Singlet fields: 20 charged, 4 neutral under all gauge symmetries
($\Phi_{0,1,3,5}$).
\item Hidden Sector:
\begin{equation}
\begin{array}{ll}
\rm T^{\{1,2,3\}}& \rm [10]\ of\ SO(10)\\
\rm D^{\{1,2,3,4,5,6,7\}}&\rm [6]\ of\ SU(4)\\
\rm \F^{\{1,2,3,4,5,6\}}& \rm [4]\ of\ SU(4)\\
\rm \Fb^{\{1,2,3,4,5,6\}}& \rm [\bar4]\ of\ SU(4)
\end{array}
\end{equation}
The $\F_i,\Fb_j$ fields carry $\pm1/2$ electric charges and exist only
\underline{confined} in hadron-like {\em cryptons} \cite{cryptons}. This is
crucial phenomenological property which is achieved naturally in stringy
flipped SU(5) \cite{ELN}.\footnote{This is to be contrasted with other
fermionic models where the confinement of such particles may be quite difficult
to achieve \cite{Alon}, and yet further models where fractionally charged
particles appear in the observable sector \cite{Lykken} and thus confinement is
not even a possibility.}
\end{itemize}
The cubic superpotential can be easily determined to be
\begin{eqnarray}
W_3=g\sqrt{2}&\biggl\{
&F_0F_1h_1+F_2F_2h_2+F_4F_4h_1+F_4\bar f_5\bar h_{45}+F_3\bar f_3\bar h_3
							\nonumber\\
            &+&\bar f_2 l^c_2 h_2+\bar f_5 l^c_5 h_2\nonumber\\
&+&\rt F_4\bar F_5\phi_3+\h F_4\bar F_4\Phi_0
                  +\bar F_4\bar F_4\bar h_1+\bar F_5\bar F_5\bar h_2\nonumber\\
&+&(h_1\bar h_2\Phi_{12}+h_2\bar h_3\Phi_{23}+h_3\bar h_1\Phi_{31}
                           +h_3\bar h_{45}\bar\phi_{45}+{\rm h.c.})\nonumber\\
&+&\h(\phi_{45}\bar\phi_{45}+\phi^+\bar\phi^++\phi^-\bar\phi^-
                        +\phi_i\bar\phi_i+h_{45}\bar h_{45})\Phi_3\nonumber\\
&+&(\eta_1\bar\eta_2+\bar\eta_1\eta_2)\Phi_0
		+(\phi_3\bar\phi_4+\bar\phi_3\phi_4)\Phi_5\nonumber\\
&+&(\Phi_{12}\Phi_{23}\Phi_{31}+\Phi_{12}\phi^+\phi^-
                +\Phi_{12}\phi_i\phi_i+{\rm h.c.})\nonumber\\
&+&T_1T_1\Phi_{31}+T_3T_3\Phi_{31}\nonumber\\
&+&D_6D_6\Phi_{23}+D_1D_2\bar\Phi_{23}+D_5D_5\bar\Phi_{23}+D_7D_7\bar\Phi_{31}
							\nonumber\\
&+&D_3D_3\Phi_{31}+\h D_5D_6\Phi_0+\rt D_5D_7\bar\phi_3\nonumber\\
&+&\F_4\Fb_6\bar\Phi_{12}+\h F_3\Fb_4\Phi_0+\h F_2\Fb_5\Phi_3
                        +\F_6\Fb_4\phi^+\nonumber\\
&+&\rt \F_5\Fb_4\phi_4+\F_1\Fb_2D_5+\F_2\Fb_4l^c_2\biggr\}\nonumber
\end{eqnarray}
Higher-order terms in the superpotential are suppressed by powers of the
string scale, but are also calculable. The quartic superpotential terms are
\begin{eqnarray}
W_4&=
&F_2\bar f_2\bar h_{45}\bar\phi_4+F_3\bar F_4 D_4D_6+F_3\bar F_5 D_4
D_7\nonumber\\
&+&l^c_3\Fb_3\Fb_6D_7+l^c_5\F_2\Fb_3\bar\phi_3
                       +\F_1\Fb_3(\phi^+\bar\phi_3+\bar\phi^-\phi_3)\nonumber\\
&+&\Fb_3\Fb_5D_7\bar\phi^-+\F_2\F_5D_3\phi^-+\F_2\F_6D_3\phi_4+\F_5\Fb_1D_2D_7
\nonumber\\
&+&\F_5\Fb_2D_1D_7+\F_3\Fb_3D_3D_6+\F_4\Fb_3D_4D_7+\F_5\Fb_4D_5D_7.
                                                        \nonumber
\end{eqnarray}
where the coefficients are generically ${\cal O}(1)$ and can be calculated
explicitly \cite{KLN}. Traditionally the above results have been used to study
the phenomenology of the models, aided by a set of non-zero vacuum expectation
values for many of the singlets fields, as required by the flatness conditions
in the presence of an anomalous $\rm U_A(1)$ (see below).

\subsection{Recent developments}
Not until recently has the K\"ahler potential of realistic fermionic models
been studied in detail \cite{LNY94}. This second phase of string model building
is essential to the study of the supersymmetry-breaking spectrum that arises,
and depends crucially on the proper identification of the moduli fields in
free-fermionic models.
\subsubsection{The K\"ahler potential}
As a first step, one notes that the fields in the model organize themselves
into three sets of untwisted and twisted fields. In this model one
finds
\begin{equation}\!\!\!\!\!\!\!\!\!\!
\begin{tabular}{lcc}
\underline{Set}&\underline{Untwisted} [$\alpha^{(I)}_i\equiv
U^{(I)}$]&\underline{Twisted} [$\beta^{(I)}_i\equiv T^{(I)}$]\\
First&$\Phi_0,\Phi_1,\Phi_{23},\bar\Phi_{23}$&$F_0,F_1,F_4,\bar F_4$\\
&$h_1,\bar h_1$&$\F_3,\Fb_{1,2,4},D_{1,2,5,6}$\\
Second\qquad\qquad&$\eta_1,\bar\eta_1,\Phi_{31},\bar\Phi_{31}$
&$F_2,\bar f_2,l^c_2;\bar F_5,\bar f_5,l^c_5$\\
&$h_2,\bar h_2$&$\F_{1,5,6},\Fb_3,D_{3,7},T_{1,3}$\\
Third&$\Phi_3,\Phi_5,\Phi_{12},\bar\Phi_{12}$
&$F_3,\bar f_3,l^c_3;h_{45},\bar h_{45}$\\
&$\eta_2,\bar\eta_2$
&$\phi_{45},\bar\phi_{45},\phi^+,\bar\phi^+,\phi^-,\bar\phi^-,\phi_{3,4},
\bar\phi_{3,4}$\\
&$h_3,\bar h_3$&$\F_{2,4},\Fb_{5,6},D_4,T_2$
\end{tabular}
\label{eq:sets}
\end{equation}
These sets are easily deduced by studying some left-moving (\ie, world-sheet
supersymmetric) charge assignments. The K\"ahler potential can then be
calculated. The crucial observation is that the model possess only one modulus
field (besides the dilaton $S$) which corresponds to the all-neutral untwisted
singlet field $\Phi_1$, which belongs to the first set. One finds \cite{LN}
\begin{eqnarray}
K&=&-\ln(S+\bar S)
-\ln\left[(\tau+\bar
\tau)^2-\sum_i^{n_{U_1}}[\alpha^{(1)}_i+\bar\alpha^{(1)}_i]^2\right]\nonumber\\
&&+\sum_i^{n_{U_2}}\alpha^{(2)}_i\bar\alpha^{(2)}_i
+\sum_i^{n_{U_3}}\alpha^{(3)}_i\bar\alpha^{(3)}_i
+\sum_i^{n_{T_1}}\beta^{(1)}_i\bar\beta^{(1)}_i\nonumber\\
&&+{1\over\left[(\tau+\bar
\tau)^2-\sum_i^{n_{U_1}}[\alpha^{(1)}_i+\bar\alpha^{(1)}_i]^2\right]^{1/2}}
\left(\sum_i^{n_{T_2}}\beta^{(2)}_i\bar\beta^{(2)}_i
+\sum_i^{n_{T_3}}\beta^{(3)}_i\bar\beta^{(3)}_i\right)
\label{eq:K}
\end{eqnarray}
Note the special role played by the modulus field ($\tau$) and the other
fields in the first unstwisted set. The gravitino mass is given by
\begin{equation}
m^2_{3/2}=e^{\vev{K}}\,\vev{|W|}^2=
{\vev{|W|}^2\over \vev{(S+\bar S)(\tau+\bar\tau)^2}}
\label{eq:m3/2}
\end{equation}
where the sole source of supersymmetry breaking is $\vev{W}\not=0$. In what
follows we assume that $\vev{W}$ is a constant independent of the moduli
fields. Moreover, as it stands the gravitino mass is undetermined, since it
depends on the flat directions $S$ and $\tau$. In what follows we assume that
these are somehow fixed to $\vev{S}\sim\vev{\tau}\sim1$ (in Planck units),
as motivated from $S$ and $T$ duality considerations. To achieve a sufficiently
small value of $m_{3/2}$ then $\vev{W}$ would have to be suppressed, as occurs
for instance in gaugino condensation models. From the low-energy effective
theory point of view, $\vev{S}$ and $\vev{\tau}$ could be determined
dynamically via the no-scale mechanism \cite{noscale}. This mechanism is
destabilized by one-loop quadratically divergent contributions to the scalar
potential, but fortunately in our model (as discussed in
Sec.~\ref{sec:shiftingQ}) these can be made to vanish in a suitably chosen
vacuum.

\subsubsection{Properties of the K\"ahler potential [23]}
The most basic property of the K\"ahler potential in Eq.~(\ref{eq:K}) is that
the tree-level vacuum energy vanishes, \ie,
\begin{equation}
V_0=m^2_{3/2}\,(1 [S] +2[\tau]-3)=0
\end{equation}
where the contributions from the two moduli are indicated. Next we need to
worry about the quadratically-divergent one-loop correction \cite{OldEKN,FKZ}
\begin{equation}
{1\over32\pi^2}{\rm Str}\,{\cal M}^2 M^2_{\rm Pl}\equiv{1\over16\pi^2}\, Q_0 \,
m^2_{3/2}\,M^2_{\rm Pl}
\end{equation}
The quantity $Q_0$ can be straightforwardly calculated from the K\"ahler
potential
\begin{equation}
Q_0=n_{U_2}+n_{U_3}+n_{T_1}-n_{U_1}-d_f-3=4
\label{eq:Q0}
\end{equation}
where the $n_{U_1}=13,n_{U_2}=14,n_{U_3}=16$ and
$n_{T_1}=80,n_{T_2}=80,n_{T_3}=68$ represent the numbers of fields in each
of the untwisted and twisted sets in Eq.~(\ref{eq:sets}), and $d_f=90$ is the
dimension of the gauge group. We note that this lowest-order calculation of
$Q_0$ yields a ``small" enough value, which can be further shifted towards
zero in the presence of an anomalous $\rm U_A(1)$ factor.

Turning to the supersymmetry-breaking parameters, the goldstino field is
\begin{equation}
\tilde\eta={S+\sqrt{2}\,\tau\over\sqrt{3}}
\end{equation}
whereas the scalar masses are given by the following multiples of $m_{3/2}$
\begin{equation}
\begin{tabular}{ccrcccr}
$U^{(1)}\,$:&$\alpha^{(1)}$& $0$ &\qquad\qquad &$T^{(1)}\,$:&$\beta^{(1)}_i$&
$1$\\
$U^{(2)}\,$:&$\alpha^{(2)}_i$& $1$ & &$T^{(2)}\,$:&$\beta^{(2)}_i$& $0$\\
$U^{(3)}\,$:&$\alpha^{(3)}_i$& $1$ & &$T^{(3)}\,$:&$\beta^{(3)}_i$& $0$\\
\end{tabular}
\label{eq:ScalarMasses}
\end{equation}
The trilinear scalar couplings are universal
\begin{equation}
A=m_{3/2}
\end{equation}
and so are the (tree-level) gaugino masses
\begin{equation}
m_{1/2}=m_{3/2}
\end{equation}
\subsubsection{Flatness constraints}
Keeping supersymmetry unbroken at the string scale requires the usual
F- and D-flatness conditions to be satisfied. The solutions to these conditions
become non-trivial in the presence of an anomalous $\rm U_A(1)$ factor in the
gauge group, as is the case of all free-fermionic models known to date.
F-flatness is given by the usual condition
\begin{equation}
\VEV{\partial W\over \partial\phi_i}=0
\end{equation}
where the $\phi_i$ run over all fields in the model. D-flatness involves
the usual conditions for the non-anomalous U(1) factors in the gauge group
\begin{equation}
\vev{D_a}=\sum_i q^i_a\vev{\phi}^2=0
\end{equation}
and the modified relation for $\rm U_A(1)$
\begin{equation}
\vev{D_A}=\sum_i q^i_A\vev{\phi}^2+ \epsilon=0\,,\quad
\epsilon={g^2M^2\over192\pi^2}\,{\rm Tr}\,{\rm U_A}>0
\end{equation}
In this expression the fundamental scale is $M\approx10^{18}\GeV$, from which
we can obtain (in this model) the new scale $\sqrt{\epsilon}$ \cite{search}
\begin{equation}
\sqrt{\epsilon}=3.7\times g\times10^{17}\GeV\sim{M\over10}
\label{eq:rooteps}
\end{equation}
The scalar fields charged under $\rm U_A(1)$ get vevs ${\cal
O}(\sqrt{\epsilon})$, such that all the F- and D-flatness constraints are
satisfied. This procedure amounts to a shift in the vacuum, from the original
one built within fermionic strings (where all vevs vanish), to a ``nearby" one
where supersymmetry is unbroken (and the vevs are ${\cal O}(\sqrt{\epsilon})$).
The ``shifted" vacuum is still a consistent string solution, but which is
not directly reachable from within fermionic strings. Nonetheless, in models
with $\rm U_A(1)$, starting from the original vacuum, one can sample a large
number of nearby vacua by the deformations that preserve flatness, which
parametrize a multi-dimensional space of allowed vevs.

The usual F- and D-flatness conditions above are sufficient as long as one
does not consider supersymmetry-breaking effects that may give masses to some
of the scalars. This is certainly the case in the model at hand, as seen in
Eq.~(\ref{eq:ScalarMasses}). If one were to consider giving a vev to a scalar
field that acquires a supersymmetry breaking mass, in the shifted vacuum the
vacuum energy will not vanish anymore. Thus, one restricts the choice of
fields to be shifted to those that do not acquire supersymmetry-breaking
masses \cite{LN}. This is a new constraint in string model-building, which
leads to more restrictive sets of allowed solutions for the shifted vevs.

\subsubsection{Shifting $Q$ [26]}
\label{sec:shiftingQ}
In the process of going to the shifted vacuum, the supersymmetry-breaking
masses that contribute to ${\rm Str}\,{\cal M}^2$ may be shifted as well,
leading to a shift in $Q_0$. One can show that in our model the ensuing
shift always lowers the value of $Q_0$ (more details are given below)
\begin{equation}
Q=Q_0-({\rm specific\ sum\ of\ vevs\ squared})
\end{equation}
One is then led to the question:
Can one find a set of vevs such that F-flatness is Ok, D-flatness in the
presence of $\rm U_A(1)$ is Ok, and $Q$ is shifted to zero? The answer is
yes, several such solutions have been found explicitly \cite{constraints}.
Moreover, the explicit solutions found have been shown to be rather typical,
demonstrating the wide applicability of this mechanism \cite{StrM^2}.

We should remark that this mechanism to shift $Q$ towards zero works only
if $V_0=0$ and $Q_0$ is ``small" to begin with. In our specific example
$Q_0$ must be positive too. If $V_0\not=0$ then in the expression for $Q_0$
(c.f. Eq.~(\ref{eq:Q0})) the relative signs of the various terms do not
cooperate and invariably one obtains large values of $Q_0$. In fact, the
model discussed here is the only known free-fermionic model where $V_0=0$,
and it is the only known free-fermionic model where $Q_0$ is ``small" and
therefore can be shifted to zero by our mechanism. In all other known models
(see Table~\ref{Table3}) $V_0\not=0$ and $|Q_0|={\cal O}(100)$. It is also
worth remarking that being able to {\em calculate} $V_0$ and $Q$ is in itself
a non-trivial matter, which has not really been done in many models so far,
perhaps because of the yet-to-be-calculated form of the K\"ahler
potential.
\begin{table}[h]
\caption{Compilation of $V_0$ and $Q_0$ values in known fermionic string
models.}
\label{Table3}
\begin{center}
\begin{tabular}{|l|r|r|}\hline
Model&$V_0$&$Q_0\ $\\ \hline
PZ \cite{PZ}&$0$&$-272$\\
ALR \cite{ALR}&$3$&$-269$\\
F \cite{Alon}&$-2$&$168$\\
AEHN \cite{revamp}&$1$&$-83$\\
LNY \cite{search}&$0$&$4$\\ \hline
\end{tabular}
\end{center}
\vspace{-0.6cm}
\end{table}

Having demonstrated the viability of models with vanishing values of $Q$,
the question remains whether higher-order contributions to the scalar
potential may generate new quadratically divergent contributions, as pointed
out in Ref.~\cite{Bagger}. In other words, is there any reason to believe that,
taking into account the full string theory, these higher-order quadratic
divergences somehow vanish automatically, as long as $Q$ vanishes? A would-be
analogous situation occurs with the string loop corrections to the gauge
kinetic function, which vanish at two and higher loops. In this case
a modular anomaly is at play. In our case we would like to conjecture that
the anomalous $\rm U_A(1)$ plays the role of the anomaly at hand and, as with
all such anomalies, as long as one takes care of their appearance at one-loop
order, higher orders are dealt with automatically. Detailed string calculations
appear needed to verify or disprove such conjecture.

\subsubsection{Specific example of $\rm U_A(1)$ and $Q$ cancellation [27]}
As discussed above, in order to demonstrate the possibility of achieving F- and
D-flatness, in the presence of supersymmetry-breaking masses, we need to impose
\begin{eqnarray}
\alpha^{(2)}&:&\VEV{\eta_1,\bar\eta_1,\Phi_{31},\bar\Phi_{31}}=0
\nonumber\\
\alpha^{(3)}&:&\VEV{\eta_2,\bar\eta_2,\Phi_{12},\bar\Phi_{12},\Phi_3,\Phi_5}
=0\label{eq:new} \\
\beta^{(1)}&:&\VEV{\nu^c_0,\nu^c_1,\nu^c_4,\nu^c_{\bar4}}=0\nonumber
\end{eqnarray}
so that the vacuum energy remains zero in the shifted vacuum. The usual F-
and D-flatness conditions become more restrictive and amount to the following
set of constraints (numbers in square brackets refer to the original set of
constraints in Ref.~\cite{search}, where the new constraints in
Eq.~(\ref{eq:new})
were not imposed)
\begin{eqnarray}
&&[6.7]\quad \left\{
\begin{array}{l}
x_{45}={1\over15}\epsilon-{1\over2}V^2_3\\
x_+-x_-={1\over15}\epsilon+{1\over2}V^2_3\\
x_{23}=-{1\over5}\epsilon\\
x_3+x_4+2x_+={2\over5}\epsilon+V^2_3
\end{array}
\right.
\nonumber\\
&&[6.8]\quad \left\{ V^2=V^2_2+V^2_3=\bar V^2_5=\bar V^2\right.
\nonumber\\
&&[6.9]\quad \left\{
\begin{array}{l}
\vev{\phi^+\phi^-+\phi_4\phi_4}=0\\
\vev{\bar\phi^+\bar\phi^-+\bar\phi_3\bar\phi_3+\bar\phi_4\bar\phi_4}=0\\
\vev{\phi_{45}\bar\phi_{45}+\phi_4\bar\phi_4+\phi^+\bar\phi^++\phi^-\bar\phi^-}
=0\\
\vev{\bar\phi_3\phi_4}=0\\
\vev{\phi_3}=0
\end{array}
\right.
\nonumber
\end{eqnarray}
where $x_i=|\vev{\varphi_i}|^2-|\vev{\bar\varphi_i}|^2$ and
$V_i=\vev{\nu^c_i}$. It also follows that $Q$ is shifted in the shifted
vacuum \cite{StrM^2}
\begin{equation}
Q=4-\coeff{1}{3}(14X+5Y)
\end{equation}
where
\begin{equation}
X=\sum_{\alpha^{(1)}}
{\vev{\alpha^{(1)}+\bar\alpha^{(1)}}^2\over\vev{\tau+\bar\tau}^2}\ ,\qquad
Y=
\sum_{\beta^{(2)}} {\vev{\beta^{(2)}\bar\beta^{(2)}}\over\vev{\tau+\bar\tau}}+
\sum_{\beta^{(3)}} {\vev{\beta^{(3)}\bar\beta^{(3)}}\over\vev{\tau+\bar\tau}}
\end{equation}
Note that $X>0$ and $Y>0$, and therefore $Q$ is always shifted towards zero.
It is important to note that this expression for $Q$ in terms of the $X,Y$
variables has been obtained in the reasonable approximation that
$\vev{\tau}\sim1$ (as assumed after Eq.~(\ref{eq:m3/2})) and
$\vev{\phi}\ll\vev{\tau}$, as would be expected from anomalous $\rm U_A(1)$
considerations. As such, large shifts in $Q$ are by construction not possible.
Also, the duality properties of the K\"ahler potential may not manifest in
the expression for $Q$ because of the approximation made. The possible non-zero
contributors to $X,Y$ are
\begin{eqnarray}
\alpha^{(1)}&:&\VEV{\Phi_0,\Phi_{23},\bar\Phi_{23}}\nonumber\\
\beta^{(2)}&:&V_2,\bar V_5\\
\beta^{(3)}&:&V_3,\VEV{\phi_{45},\bar\phi_{45},\phi^+,\bar\phi^+,\phi^-,
\bar\phi^-,\phi_{3,4},\bar\phi_{3,4}}\nonumber
\end{eqnarray}
Scanning over the multi-dimensional parameter space of vevs one can find
solutions to the F- and D-flatness conditions (respecting the new constraints
in Eq.~(\ref{eq:new})). Sample values are shown in Table~\ref{Table4}. One
can then see whether these solutions also lead to $Q=0$. This is the case for
several of them (the ones with the larger values of $\sum\widehat\beta^2$), as
$Q$ also depends on vevs (in the $X$ contribution) which are not too
constrained by the flatness conditions, and thus can be adjusted to yield
$Q=0$.

\begin{table}[t]
\caption{Sample solutions to the F- and D-flatness constraints, respecting the
new constraints from $V_0=0$ in the shifted vacuum. All vevs in units of
$\epsilon^{1/2}$.}
\label{Table4}
\begin{center}
\begin{tabular}{ccccccccc|c}
$\widehat\phi_{45}$&$\widehat{\bar\phi_{45}}$&$\widehat\phi^+$
&$\widehat{\bar\phi^+}$&$\widehat\phi^-$&$\widehat{\bar\phi^-}$&
$\widehat\phi_4$&$\widehat{\bar\phi_4}$&$\widehat V_3$&$\sum\widehat\beta^2$\\
\hline
0.836&-0.796&0.531&0.368&-0.539&-0.472&0.535&0.417&0.060&2.73\\
0.999&-1.072&0.942&0.741&-0.386&-0.279&0.603&0.455&0.659&5.25\\
0.275&-0.132&0.401&0.089&-0.256&-0.008&0.320&0.026&0.130&0.46\\
0.589&-0.702&0.675&0.341&-0.298&-0.213&0.448&0.269&0.650&2.66\\
0.735&-0.910&0.892&0.559&-0.265&-0.150&0.486&0.290&0.842&4.31\\
0.835&-0.837&0.376&0.051&-0.750&-0.762&0.531&0.197&0.374&3.28\\
0.579&-0.545&0.401&0.074&-0.467&-0.416&0.433&0.175&0.236&1.52\\
0.927&-0.987&0.829&0.613&-0.435&-0.344&0.600&0.459&0.603&4.50\\
\hline
\end{tabular}
\end{center}
\end{table}

\subsection{Particle masses}
Superparticle masses receive contributions from two sources: from the masses of
their Standard Model partners and from supersymmetry breaking. Experimentally
we now know that the latter contribution dominates (except perhaps in the
case of a light top-squark). These supersymmetry-breaking contributions have
been given above (see Eq.~(\ref{eq:ScalarMasses})) and scale with $m_{3/2}$.
On the other hand, particle masses (such as quarks and leptons) can be
determined from three inputs:
\begin{itemize}
\item A mass scale, which must be $M_Z$ as masses are protected (\ie, vanish)
above the scale of SU(2)$\times$U(1) breaking
\item A Yukawa coupling, which is a pure number calculable only in string
models, where one typically obtains
\begin{equation}
\lambda\sim g\quad {\rm (gauge\ coupling)}
\end{equation}
Note that GUTs can \underline{relate} but not predict Yukawa couplings.
\item A dynamical coefficient coming from the evolution of the Yukawa
couplings from the string scale down to the electroweak scale
($\lambda(M_{Pl})\to \lambda(M_Z)$), and from low-energy mixing effects
(\eg, $\tan\beta$).
\end{itemize}
Given this scenario one would then expect $m_q\sim \lambda M_Z$, or in somebody
else's words
\begin{center}
``Ask not why the top-quark is so heavy---ask why the other quarks are so
light."
\end{center}
Indeed, in fermionic models one has a scheme to try to explain the hierarchy
of fermion masses, as non-zero Yukawa couplings are rather restricted by the
many stringy selection rules at play \cite{HFM,decisive}. Typically at the
cubic level only some
fermions acquire Yukawa couplings (\eg, those of the third generation). Yukawa
couplings for the lighter generations appear at higher-order in
non-renormalizable interactions,
\begin{equation}
\begin{array}{lll}
\lambda Q_3 t^c H&\quad& \lambda_t\sim g\\
\lambda Q_2 c^c H {\vev{\phi}\over M}&&\lambda_c\sim g{\vev{\phi}\over M}\\
\lambda Q_1 u^c H {\vev{\phi}^2\over M^2}&&\lambda_u\sim g{\vev{\phi}^2\over
M^2}\\
\end{array}
\end{equation}
A hierarchy of effective Yukawa couplings is generated once the ratios
$\vev{\phi}/M\sim1/10$ (from $\rm U_A(1)$ cancellation) are inserted. In
practice this exercise is complicated by the question of how to embed the
Standard Model fields in the string representations. Moreover, the resulting
Yukawa matrix need not be diagonal. Taking all these effects into account it
is possible to obtain semi-realistic fermion mass spectra \cite{decisive,Alon}.

\begin{figure}[t]
\vspace{4.5in}
\includegraphics{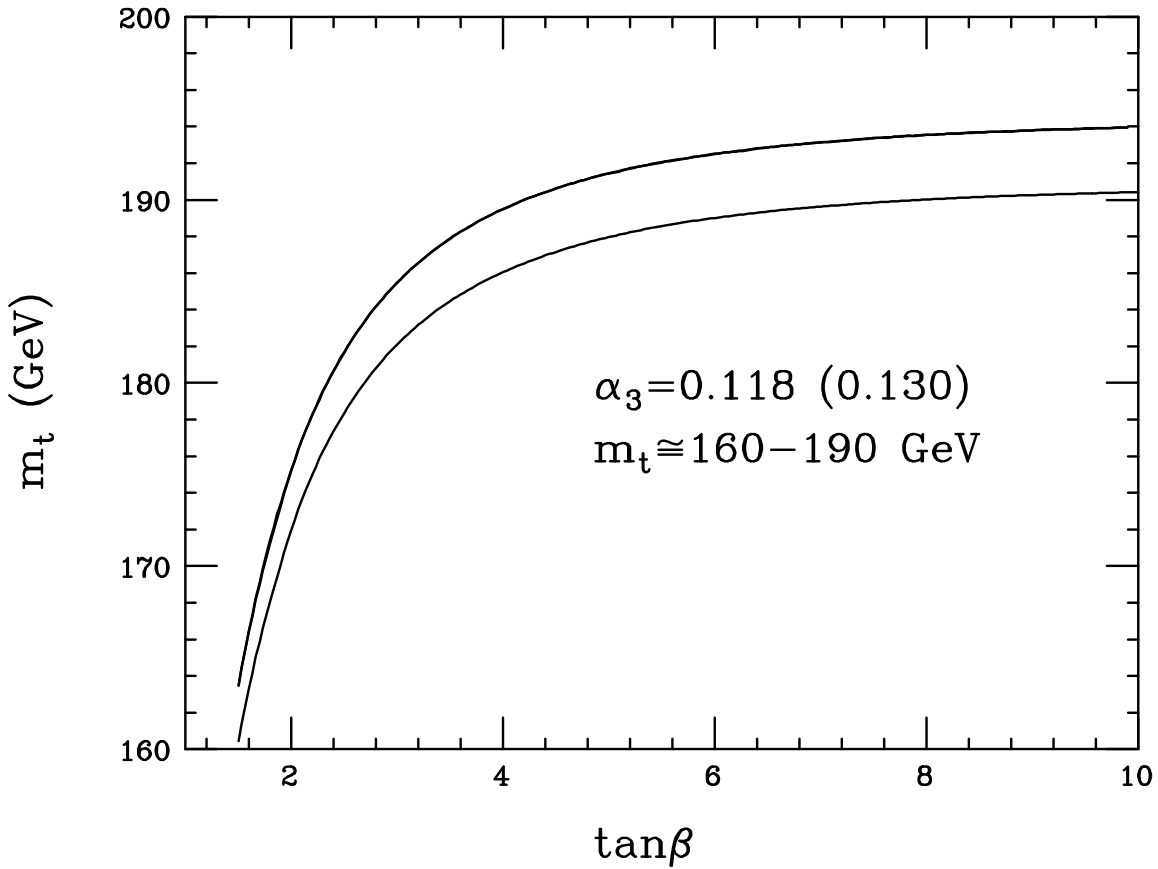}
\vspace{-0.2cm}
\caption{The calculated value of the top-quark mass as a function of
$\tan\beta$ for $\alpha_3(M_Z)=0.118\,(0.130)$ (lower (upper) curve).}
\label{fig:t-zeroGeneral}
\end{figure}

By far the simplest Yukawa coupling to calculate is that of the top quark
\begin{equation}
\lambda_t(M_{Pl})=(g\sqrt{2})\left({g\over\sqrt{2}}\right)=g^2\approx0.7
\end{equation}
where we have inserted a normalization factor ($g/\sqrt{2}$) coming from the
K\"ahler function normalization \cite{LN}, and the value of $g$ from running
the Standard Model gauge couplings up to the string scale. The top-quark mass
itself,
\begin{equation}
 m_t=\left({v\over\sqrt{2}}\right)\lambda_t(m_t)
{\tan\beta\over\sqrt{\tan^2\beta+1}}\ ,
\label{eq:mt}
\end{equation}
involves dynamics: running from $M_{Pl}\to M_Z$ to obtain $\lambda_t(M_Z)$, and
at $M_Z$ the ratio of Higgs vevs ($\tan\beta$). (There is also a $+7\%$ QCD
correction to the running mass to obtain the experimentally observable ``pole"
mass.) The result for $m_t$ as a function of $\tan\beta$ is shown in
Fig.~\ref{fig:t-zeroGeneral}, from where we conclude that
\begin{equation}
m_t\approx(160-190)\GeV\ ,
\label{eq:mtrange}
\end{equation}
a result in good agreement with experimental observations. Such
first-principles predictions should not be confused with similar numerical
results obtained in SO(10) GUT models, where the values of the $b$ and $\tau$
masses are used to deduce $m_t$.
%\clearpage

\subsection{String unification: a new scenario}
\label{sec:unification}
Typical string models possess gauge groups which are products of various
factors. In fact, all of the corresponding gauge couplings are predicted
to be equal \cite{Ginsparg} at the string scale \cite{Kaplunovsky}
\begin{equation}
M_{\rm string}=5\times g\times10^{17}\GeV\ .
\label{eq:Mstring}
\end{equation}
(This statement holds for gauge groups realized with level-one Kac-Moody
algebras, as in the case in our model of interest.) The crucial point is
that the defining property of GUTs, namely the unification of the gauge
couplings at and above the GUT scale, is obtained in string models without
appealing to traditional GUT structures. These models have then been called
Grand Unified Superstring Theories or GUSTs \cite{ELN}.

In the case of flipped SU(5), two scenarios for string unification may be
considered, depending on where the SU(5)$\times$U(1) gauge symmetry breaks
down to the Standard Model gauge group. In either scenario, the
\underline{mechanism} for generating the GUT-symmetry breaking vevs is the
vacuum shift that results in the cancellation of the anomalous $\rm U_A(1)$. In
the original scenario \cite{LNZI}, one assumed that these vevs were as large as
possible consistent with this cancellation mechanism, with the GUT symmetry
breaking occuring at the string scale.
This scenario appears quite reasonable as the string unification scale
(\ref{eq:Mstring}) nearly coincides with the scale from $\rm U_A(1)$
cancellation (\ref{eq:rooteps}), which is assumed to provide the
SU(5)$\times$U(1)-breaking vevs. Recent work considers a
slightly different scenario \cite{homeotic}, wherein SU(5)$\times$U(1) breaks
near the traditional GUT scale ($M_{32}$) and SU(5) and U(1) eventually
``superunify" at the string scale. This scale is more non-trivial to derive,
and our mechanism plays a more unique role. Such range of vevs are
perfectly allowed, and essentially determine the flat direction along which
SU(5)$\times$U(1) breaks.\footnote{It is interesting to point out that in
attempts at constructing traditional GUT models in strings (via higher-level
Kac-Moody algebras) \cite{k>1}, the generation of the GUT scale has remained
unclear. If such models ever prove to be realistic, and contain an anomalous
$\rm U_A(1)$ factor, then our mechanism should offer a possibility to obtain
the GUT scale.} The new scenario may also be more appealing to some, as
$M_{32}\ll M_{\rm string}$, making effective field theory calculations more
reliable.

The recent ``two-step" scenario has some interesting consequences not present
in the original ``one-step" scenario. First of all, in either scenario
one must consider string models with an extra pair of (\r{10},\rb{10}) SU(5)
representations to possibly achieve string unification. (This statement has
been recently re-examined in Ref.~\cite{DF}, and shown to be unavoidable.)
In the two-step scenario the (\r{10},\rb{10}) mass scale ($M_{10}$) needs to be
\cite{homeotic}
\begin{equation}
 M_{10}\sim10^{8-9}\GeV\ .
\end{equation}
On the other hand, in the one-step scenario one must split the $Q,\bar Q$ from
the $D^c,\bar D^c$ pieces in the \r{10},\rb{10}, giving them masses of order
$10^{12}\GeV$ and $10^6\GeV$ respectively \cite{LNZI}. In either scenario these
scales are in principle calculable from the scales of hidden matter
condensation, although in the two-step scenario one needs to generate only one
such scale. The breaking of SU(5)$\times$U(1) is assumed to occur in both
scenarios in the shifting vacuum process; in the two-step scenario this
requires a smaller vev, which is certainly possible (see first row in
Table~\ref{Table4}). The two-step scenario also makes use of the apparent
``LEP" scale ($M_{\rm LEP}\approx10^{16}\GeV$), since
SU(3) and SU(2) unify at this scale irrespective of the presence of the extra
(\r{10},\rb{10}), as complete SU(5) multiplets do not affect this
result.\footnote{Note that if one wants to obtain the apparent LEP scale in a
string model, SU(3) and SU(2) need to be unified at this scale, a result not
possible in either standard-like SU(3)$\times$SU(2)$\times$U(1)
\cite{FNY,Alon}, or Pati-Salam SU(4)$\times$SU(2)$\times$SU(2) \cite{ALR}, or
SU(3)$^3$ \cite{Ross} type string models, as these string-unify SU(3) and
SU(2) at $M_{\rm string}\gg M_{\rm LEP}$.}

\begin{figure}[p]
\vspace{5in}
\includegraphics{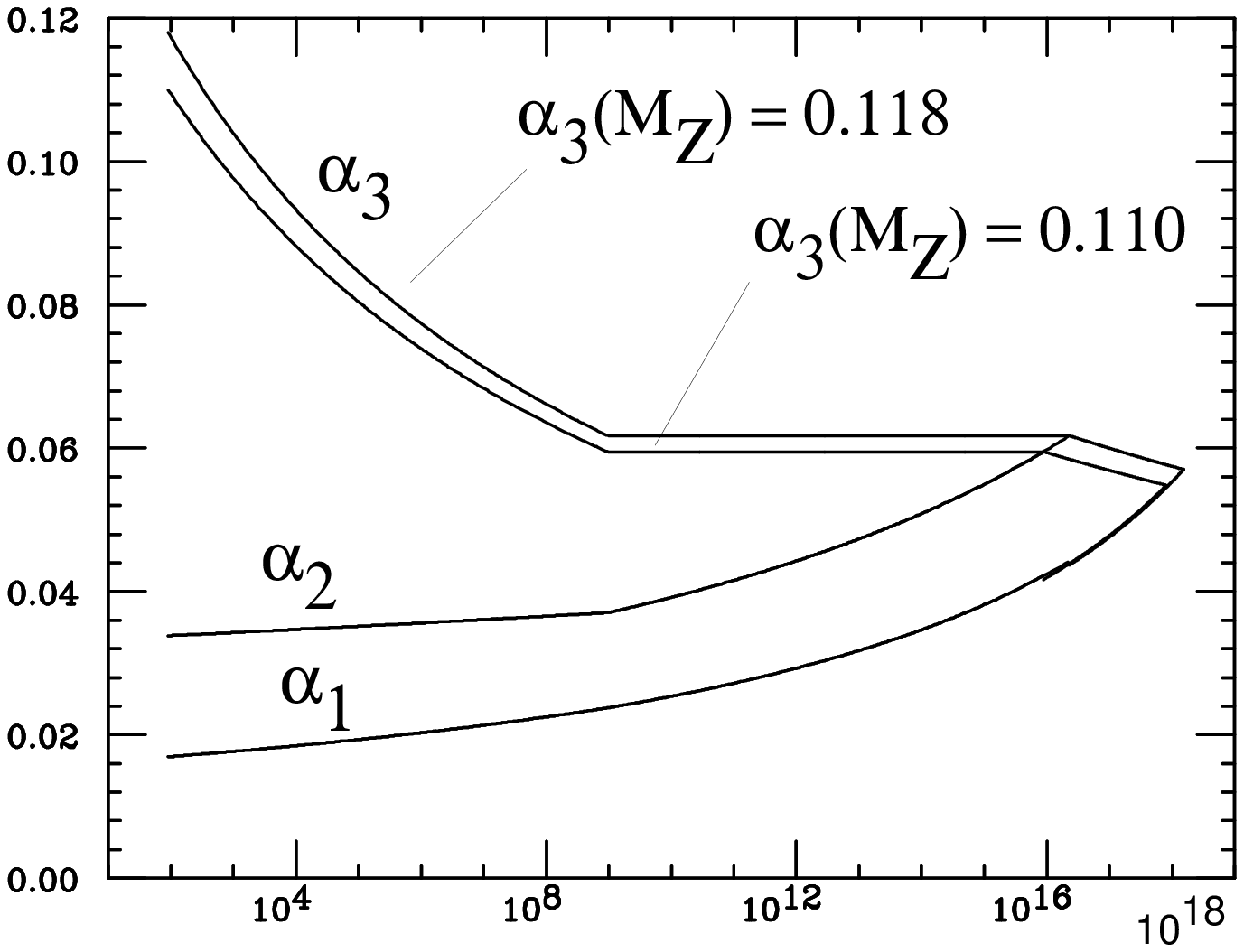}
\vspace{4.in}
\includegraphics{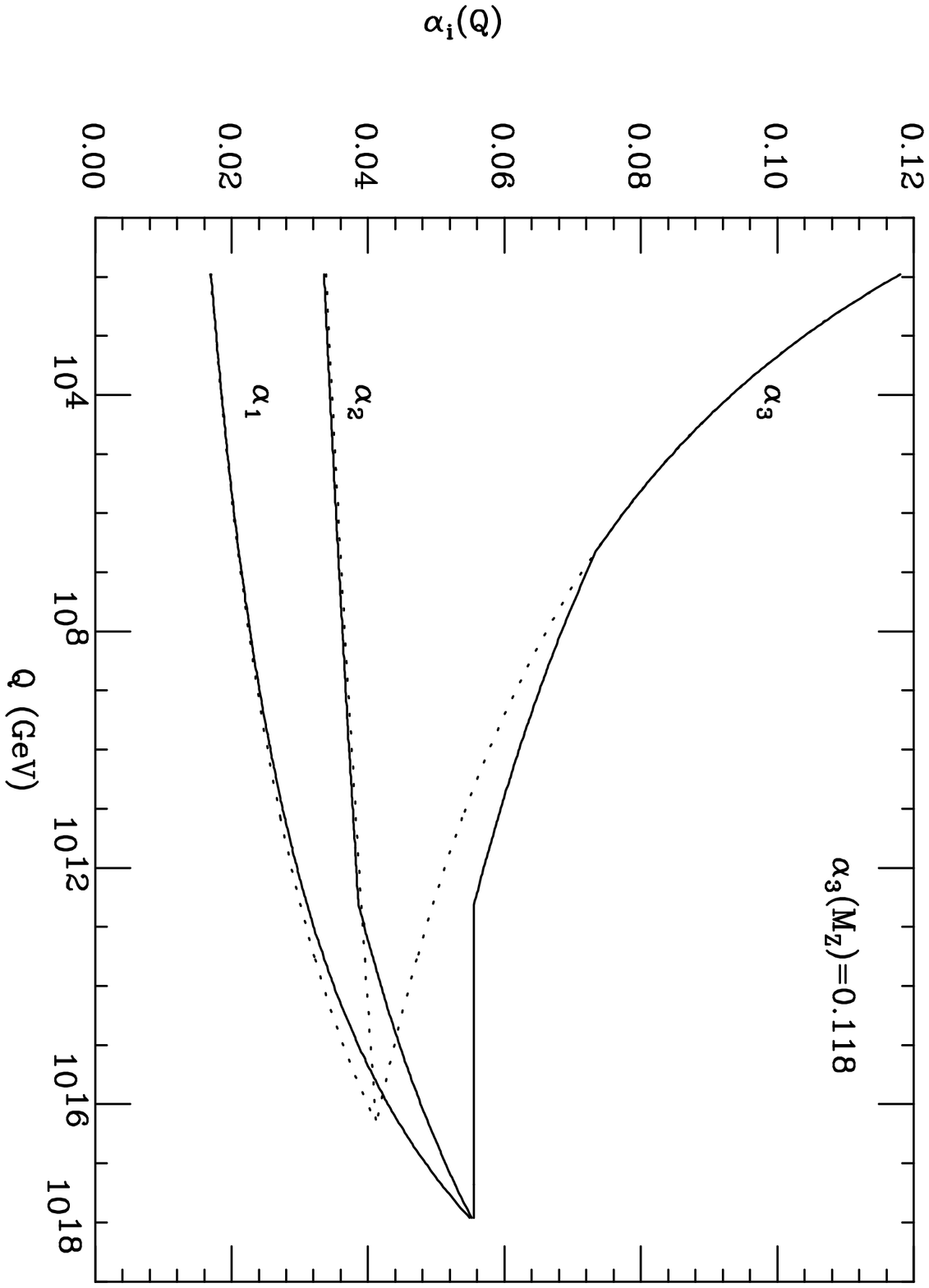}
\vspace{-2.5cm}
\caption{Two-step and one-step scenarios for string unification in flipped
SU(5).}
\label{fig:scenarios}
\end{figure}

The running of the Standard Model gauge couplings in both two- and one-step
scenarios are shown in Fig.~\ref{fig:scenarios}. The dotted lines in the
one-step scenario correspond to the case of minimal SU(5) unification. Note
that to lowest order, in the two-step scenario, $M_{32}\sim M_{\rm LEP}$ and
the SU(3) beta function vanishes above the $M_{10}$ scale. Also, in the actual
string model one expects the appearance of fractionally charged particles in
the hidden sector which are likely to modify the running of the gauge
couplings. For our present purposes we have assumed that all these particles
acquire masses of the order $\sqrt{\epsilon}\approx M_{\rm string}$, and thus
do not contribute to the gauge coupling RGEs. In a full analysis such effects
will have to be consistently included.

\section{Experimental Flipped SU(5)}
\label{sec:experimental}
As we have seen above, string models are quite predictive, with for instance
all superparticle masses ($\propto m_{3/2}$) and fermion masses calculable.
However, one should realize that predictions do vary from model to model.
Moreover, string models are complicated: one must choose one out of many-many
possible models, they have hidden and observable sectors, they have unwanted
fields that must decouple, the various intermediate scales need to be
generated dynamically, etc. Nonetheless, models can be worked out and
predictions can be derived. In this section we would like to describe a
possible scenario for a flipped SU(5) string model, as pertains to the
superparticle masses. We warn the reader that other scenarios may
exist, depending on how one assigns the Standard Model particles to the
string representations. \underline{All} of these possible scenarios are
expected to yield predictions for the top-quark mass within the range in
Eq.~(\ref{eq:mtrange}), although the required value of $\tan\beta$ and the
precise $m_t$ prediction may differ from scenario to scenario.

\subsection{A possible scenario [40]}
In this scenario one obtains for the supersymmetry-breaking masses
\begin{itemize}
\item Gaugino masses (universal)
\begin{equation}
m_{1/2}=m_{3/2}
\label{eq:susy-first}
\end{equation}
\item Scalar masses
\begin{eqnarray}
{\rm First\ generation}&&m^2_{Q_1,U_1^c,D_1^c,L_1,E_1^c}=0\\
{\rm Second\ generation}&&m^2_{Q_2,U_2^c,D_2^c,L_2,E_2^c}=0\\
{\rm Third\ generation}&&m^2_{Q_3,D^c_3}=m^2_{3/2},\ m^2_{U^c_3,L_3,E^c_3}=0\\
{\rm Higgs\ masses}&&m^2_{H_1}=m^2_{H_2}=0
\end{eqnarray}
\item Trilinear scalar couplings (universal)
\begin{equation}
A_0=m_{3/2}
\end{equation}
\item Bilinear scalar coupling
\begin{equation}
B_0=m_{3/2}
\label{eq:susy-last}
\end{equation}
\end{itemize}
Our scenario has only one free parameter. At high energies our unknowns are
$m_{3/2}$ and $\mu$.\footnote{In principle, even $\mu$ is calculable in this
class of models. We expect to obtain an effective $\mu$ term from
non-renormalizable contributions to the superpotential, \ie, $\lambda h\bar
h\vev{TT}/M\Rightarrow\mu=\lambda\vev{TT}/M$.} At low-energies we inherit
the ratio of Higgs vacuum expectation values ($\tan\beta$), which must exceed
unity for the radiative electroweak breaking mechanism to work. This
mechanism imposes two constraints on the model parameters, which can be used
to determine $|\mu|$ and $\tan\beta$ in terms of $m_{3/2}$. Thus, our model
can be described in terms of a single parameter:
$m_{1/2}\leftrightarrow m_{\tilde g}\leftrightarrow m_{\chi^\pm_1}$

To obtain the low-energy spectrum, starting from the high-energy
supersymmetry-breaking mass parameters in
Eqs.~(\ref{eq:susy-first}-\ref{eq:susy-last}), we need to follow the standard
procedure of running the coupled renormalization group equations for all
masses, gauge, and Yukawa couplings. The following analysis has been carried
out in the one-step scenario for string unification, as discussed in
Sec.~\ref{sec:unification}. Calculations for the new two-step scenario are
underway \cite{homeotic}. At the electroweak scale we minimize the
one-loop effective potential and determine $|\mu|$ and $B$ (see \eg,
Ref.~\cite{aspects}). We  adjust $\tan\beta$ until the computed $B_0$ agrees
with first-principles calculation of $B_0$ in Eq.~(\ref{eq:susy-last}). One
finds that only
\begin{equation}
\mu<0
\end{equation}
works, and that the values of $\tan\beta$, even though $m_{3/2}$ dependent,
fall in the following narrow range
\begin{equation}
\tan\beta\approx2.2-2.3\ .
\end{equation}
Therefore, from Eq.~(\ref{eq:mt}) one can determine $m_t$
\begin{equation}
m_t\approx175\,(179)\GeV,\quad \alpha_3=0.118\,(0.130)\ .
\end{equation}
\begin{figure}[t]
\vspace{4.5in}
\includegraphics{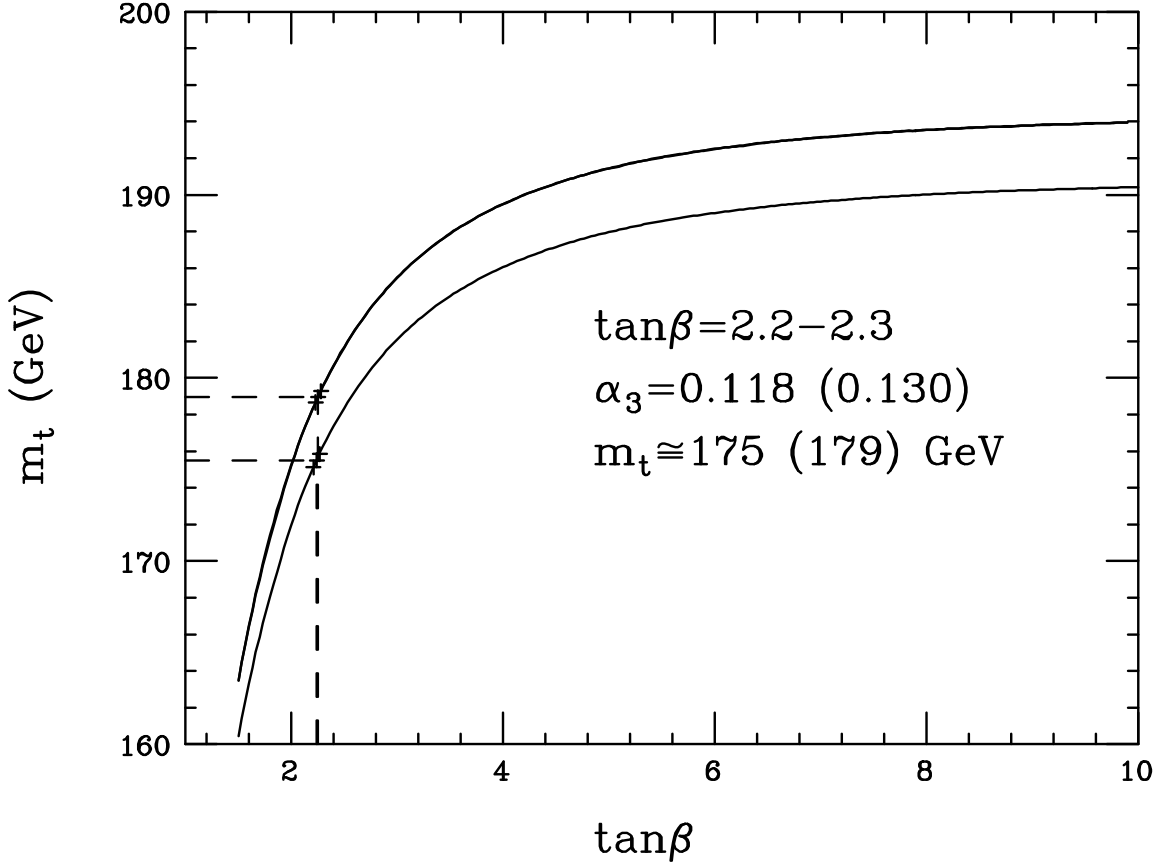}
\vspace{-0.2cm}
\caption{The calculated value of the top-quark mass as a function of
$\tan\beta$ for $\alpha_3(M_Z)=0.118\,(0.130)$ (lower (upper) curve),
indicating the predicted value of $\tan\beta$, and therefore of the top-quark
mass.}
\label{fig:t-zeroR}
\end{figure}
This result is shown in Fig.~\ref{fig:t-zeroR}.

\begin{figure}[p]
\vspace{6in}
\includegraphics{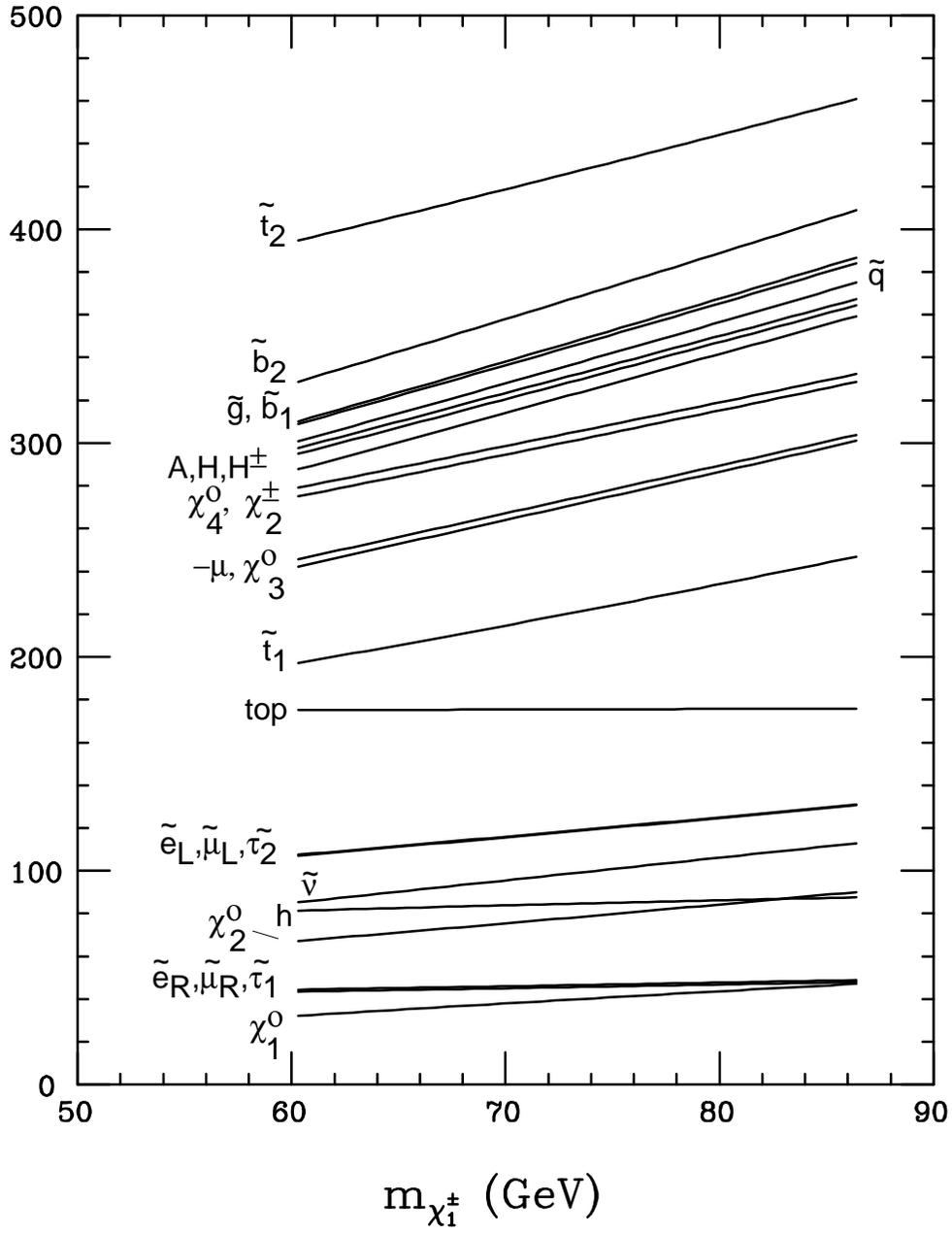}
\vspace{2cm}
\caption{Full spectrum of superparticles in a possible scenario of stringy
flipped SU(5). Note the cutoff in the parameter space resulting from demanding
a neutral and colorless lightest supersymmetric particle (LSP).}
\label{fig:spectrum}
\end{figure}

The determination of the full spectrum of superparticle masses then follows,
although not without subtlety. Because the non-universality of scalar masses at
the string scale, the renormalization group equations receive a new
contribution, which effectively amounts to a shift in all scalar mass
parameters \cite{LM}
\begin{equation}
\Delta m^2_i=-c^2Y_if\ ,
\end{equation}
where $Y_i$ is the hypercharge of the particle, $f=0.0060$ is an RGE
factor, and $c^2$ is the non-universality coefficient at the string scale
\begin{equation}
c^2=m^2_{H_2}-m^2_{H_1}+\sum_{i=1,2,3}\left(m^2_{Q_i}+m^2_{D^c_i}+m^2_{E^c_i}
-m^2_{L_i}-2m^2_{U^c_i}\right)\ .
\end{equation}
In our case $c^2=2m^2_{1/2}$. This shift affects all sparticle masses, but
is significant only for the right-handed sleptons, as these possess only
hypercharge quantum numbers. The right-handed slepton masses are given by
\begin{equation}
m_{\tilde\ell_R}^2=a\,m^2_{1/2}+\tan^2\theta_W
M^2_W{\tan^2\beta-1\over\tan^2\beta+1}
\end{equation}
with $a=0.153$ in the usual universal case, but $a=0.153-0.120=0.034$ here.
This mass should be compared with that of the lightest neutralino $\chi^0_1$
\begin{eqnarray}
m_{\chi^0_1}&\approx&0.25 m_{1/2}\\
m_{\tilde\ell_R}&\approx&\sqrt{(0.18m_{1/2})^2+(36)^2}\ ,
\end{eqnarray}
which shows that if $m_{1/2}$ is too large, then
$\tilde\ell_R$ becomes the lightest supersymmetric particle (LSP). Since such
electrically charged particle would be stable, our model cannot support such
regime, and we obtain an important cutoff in the one-dimensional parameter
space
\begin{equation}
m_{1/2}\lsim180\GeV\ .
\end{equation}
The full spectrum is shown in Fig.~\ref{fig:spectrum}. Of note are the
following features
\begin{eqnarray}
&m_{\chi^\pm_1}\approx m_{\chi^0_2}<90\GeV\\
&m_h<90\GeV\\
&m_{\tilde\ell_R}<50\GeV\\
&m_{\tilde q}\approx 0.98m_{\tilde g}\\
&\tilde t_{1,2}\ {\rm big\ split}
\end{eqnarray}

\subsection{Discovery prospects at the Tevatron}
The spectrum of the model, as shown in Fig.~\ref{fig:spectrum} indicates that
the traditional supersymmetry signals at hadron colliders, namely searches
for squarks and gluinos via the missing-energy channel, are not accessible
at the present-day Tevatron, as one has $m_{\tilde q,\tilde
g}\approx(300-400)\GeV$. Some of this parameter space may become accessible
in the Main Injector era, and more so in the case of a high-luminosity upgrade
of the Tevatron \cite{DiTevatron}. This mass range would be, however, easily
accessible at the LHC.

\begin{figure}[t]
\vspace{4.6in}
\includegraphics{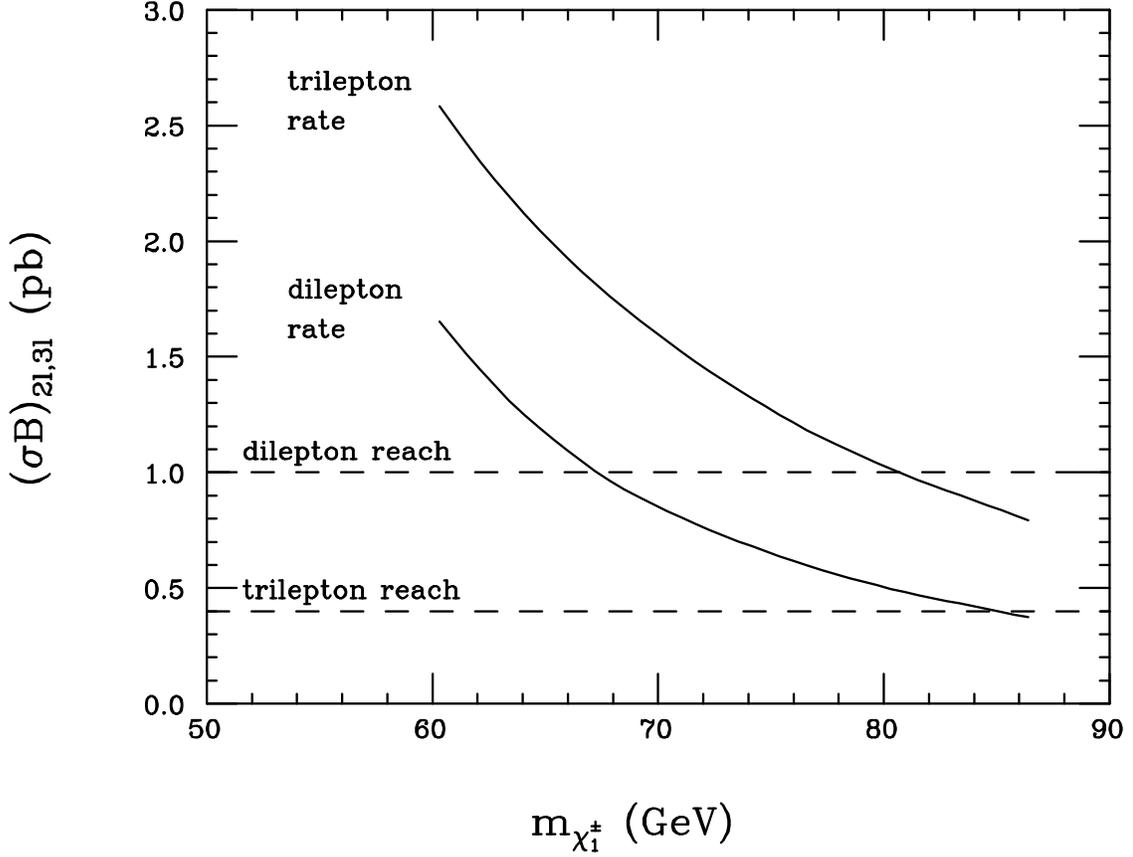}
\vspace{0.2cm}
\caption{Calculated trilepton and dilepton rates from chargino-neutralino
production at the Tevatron versus the chargino mass. Also indicated are the
expected experimental sensitivities with $100\,{\rm pb}^{-1}$ of accumulated
data.}
\label{fig:di-tri}
\end{figure}

A much more promising and immediate avenue for discovery is afforded by
considering the production and decay of neutralinos and charginos, as the
lightest of these must satisfy $m_{\chi^0_2,\chi^\pm_1}<90\GeV$. The reactions
of interest are \cite{trileptons,di-tri}
\begin{eqnarray}
&p\bar p\to\chi^0_2\chi^\pm_1X\to3\ell\quad ({\rm trileptons})\\
&p\bar p\to\chi^+_1\chi^-_1X\to2\ell\quad ({\rm dileptons})
\end{eqnarray}
The corresponding branching ratios can be calculated, as everything else, in
terms of our one parameter, giving \cite{zero}
\begin{eqnarray}
B(\chi^\pm_1\to \ell^\pm)&\approx&1/2\quad (\ell=e+\mu)\\
B(\chi^\pm_1\to2j)&\approx&1/4\\
B(\chi^0_2\to \ell^+\ell^-)&=&2/3
\end{eqnarray}
(Note that the last ratio is maximal, as this channel proceeds via the two-body
decay $\chi^0_2\to\tilde\ell_R\ell$). These sizeable branching ratios and
relatively light superparticles imply significant trilepton and dilepton rates.
These are shown in Fig.~\ref{fig:di-tri}, along with an estimate of the
experimental sensitivity expected by the end of Run IB ($\int{\cal
L}=100\ipb$) \cite{di-tri}. Clearly, complete exploration of this model at the
Tevatron via trilepton events is guaranteed.

Searches for the light sleptons predicted in this model
($m_{\tilde\ell_R}<50\GeV$) appear also possible via
$p\bar p\to\tilde\ell^+_R\tilde\ell^-_RX$. This process has a cross section of
$1\,(0.1)\pb$ for $m_{\tilde\ell_R}=45\,(50)\GeV$, as we expect
$B(\tilde\ell_R\to\ell\chi^0_1)=1$ into dileptons. However, the prospects
for slepton detection are not very bright, given the low rates and large
backgrounds that are expected.

\subsection{Discovery prospects at LEPII}
At LEPII, starting to operate in the Fall of 1995, one could detect the
lightest Higgs boson, the lightest chargino, and the right-handed sleptons.
Discovery of the latter two should be straightforward, with only a slight
increase in the beam energy and a modest amount of luminosity.

Charginos are predicted to obey $m_{\chi^\pm_1}\approx(60-90)\GeV$, and should
be most easily detectable via \cite{Grivaz}
\begin{equation}
e^+e^-\to \chi^+_1\chi^-_1\to \ell+2j+{\rm missing}
\end{equation}
Inserting the chargino branching ratios given above, for
$m_{\chi^\pm_1}=60\,(90)\GeV$ we expect a cross section of $2.1\,(0.62)\pb$
(at $\sqrt{s}=200\GeV$). With a sufficiently high beam energy, chargino
production into the ``mixed" mode should be quite noticeable.
Right-handed sleptons must be rather light in this model
($m_{\tilde\ell_R}<50\GeV$) and will be pair-produced
\begin{equation}
e^+e^-\to \tilde\ell^+_R\tilde\ell^-_R\to2\ell+{\rm missing}
\end{equation}
We expect $\sigma(\tilde e_R)>2\pb$ and $\sigma(\tilde\mu_R)>0.9\pb$ (at
$\sqrt{s}=200\GeV$), which should make detection immediate. Finally, because of
the prediction $\tan\beta\approx2.2$, the Higgs boson will be at the low end of
its possible range ($m_h\approx80-90\GeV$), and should be produced via
\cite{EGN}
\begin{equation}
e^+e^-\to Z^*\to Zh\to (f\bar f) (b\bar b)
\end{equation}
Such Higgs boson will be indistinguishable from the Standard Model Higgs boson
($\sin^2(\alpha-\beta)=0.996-0.998$). The estimated mass reach of LEPII as a
function of the center-of-mass energy \cite{Sopczak},
\begin{equation}
m_h\approx\sqrt{s}-95\GeV\ ,
\end{equation}
will make it sensitive to our Higgs boson for $\sqrt{s}\gsim(175-185)\GeV$.
This is not expected to occur in the initial phase of the LEPII upgrade.
It is worth remarking that the supersymmetry channel $h\to\chi^0_1\chi^0_1$
erodes the usual $h\to b\bar b$ signal a little.

\subsection{Discovery prospects via Rare processes}
We consider three rare processes through which this model could be tested
indirectly: $b\to s\gamma$ at CLEO, $\rm(g-2)_\mu$ at Brookhaven, and $R_b$ at
LEP.
\subsubsection{${\rm B}(b\to s\gamma)$}
This decay mode has been observed by the CLEO Collaboration with the following
result \cite{CLEO}
\begin{equation}
{\rm B}(b\to s\gamma)^{\rm exp}=(1-4)\times10^{-4}\ .
\label{eq:bsg-exp}
\end{equation}
This result agrees well with the Standard Model prediction (for all allowed
values of $m_t$) and therefore constrains any extension of the Standard Model.
In particular, in supersymmetry ${\rm B}(b\to s\gamma)$ varies a lot over the
multi-dimensional parameter space \cite{bsgamma}. In fact, it is possible to
find regions of parameter space where the supersymmetry prediction is larger or
smaller than the Standard Model predicition by more than two orders of
magnitude. The dominant diagram involves the chargino--top-squark loop and
depends strongly on $\tan\beta$. Also, QCD corrections are large, and to date
only fully known to leading order. In the present model we find the following
range \cite{zero}
\begin{equation}
{\rm B}(b\to s\gamma)=[(4.2-5.3)\to(3.9-5.1)]\times10^{-4}\ ,
\end{equation}
which is in fair agreement with the experimental result in
Eq.~(\ref{eq:bsg-exp}). (The uncertainties due to uncalculated next-to-leading
order QCD corrections have been estimated by varying the renormalization scale
around the $b$-quark mass \cite{Buras}.) It should be noted that our small
value of $\tan\beta$ helps in suppressing the supersymmetric contributions
sufficiently, given the lightness of our spectrum.

\subsubsection{$\rm (g-2)_\mu$}
The anomalous magnetic moment of the muon was last measured in 1970, with the
result $a^{\rm exp}_\mu=1165923\,(8.5)\times10^{-9}$ \cite{g-2-exp}. When this
result is contrasted with the present Standard Model prediction
($a^{\rm SM}_\mu=1165919.20\,(1.76)\times10^{-9}$ \cite{kinoshita}), it allows
a 95\%CL interval for new physics contributions \cite{g-2}
\begin{equation}
-13.2\times10^{-9}<a^{susy}_\mu<20.8\times10^{-9}\ .
\label{eq:g-2range}
\end{equation}
There are two supersymmetric contributions to $a_\mu$, with charginos and
sneutrinos in the loop, or with sleptons and neutralinos in the loop. The
latter is small because of the small $\tilde\mu_L-\tilde\mu_R$ mixing angle
($\propto m_\mu$). The dominant chargino-sneutrino contribution is greatly
enhanced by large values of $\tan\beta$ (not unlike the chargino--top-squark
contribution in $b\to s\gamma$), and can lead to values of $a^{\rm susy}_\mu$
outside the allowed range in Eq.~(\ref{eq:g-2range}) \cite{g-2}. Nonetheless we
find \cite{zero}
\begin{equation}
a^{\rm susy}_\mu=(-2.4\to-1.7)\times10^{-9}
\end{equation}
which is within the present limits. The new E821 experiment at Brookhaven,
scheduled to start taking date in 1996, should achieve a sensitivity of
$0.4\times10^{-9}$ \cite{Roberts} and thus be greatly sensitive to this
prediction.

\subsubsection{$R_b$}
The ratio $R_b=\Gamma(Z\to b\bar b)/\Gamma(Z\to{\rm hadrons})$ has been
measured at LEP to be $R^{\exp}_b=0.2219\pm0.0017$ \cite{Rb-exp}, which is more
than three standard deviations above the Standard Model prediction of $R_b^{\rm
SM}=0.2157$. We obtain \cite{zero}
\begin{equation}
R_b^{\rm susy}=(4.4\to3.2)\times10^{-4}\ ,
\end{equation}
which shifts the Standard Model prediction in the direction of the experimental
result only slightly. This shift will not be observable unless the
experimental sensitivity increases by a factor of four, and then only if the
experimental result has somehow been reconciled with the Standard Model
prediction.

\subsection{Prospects for Dark Matter detection}
\begin{figure}[t]
\vspace{5.5in}
\includegraphics{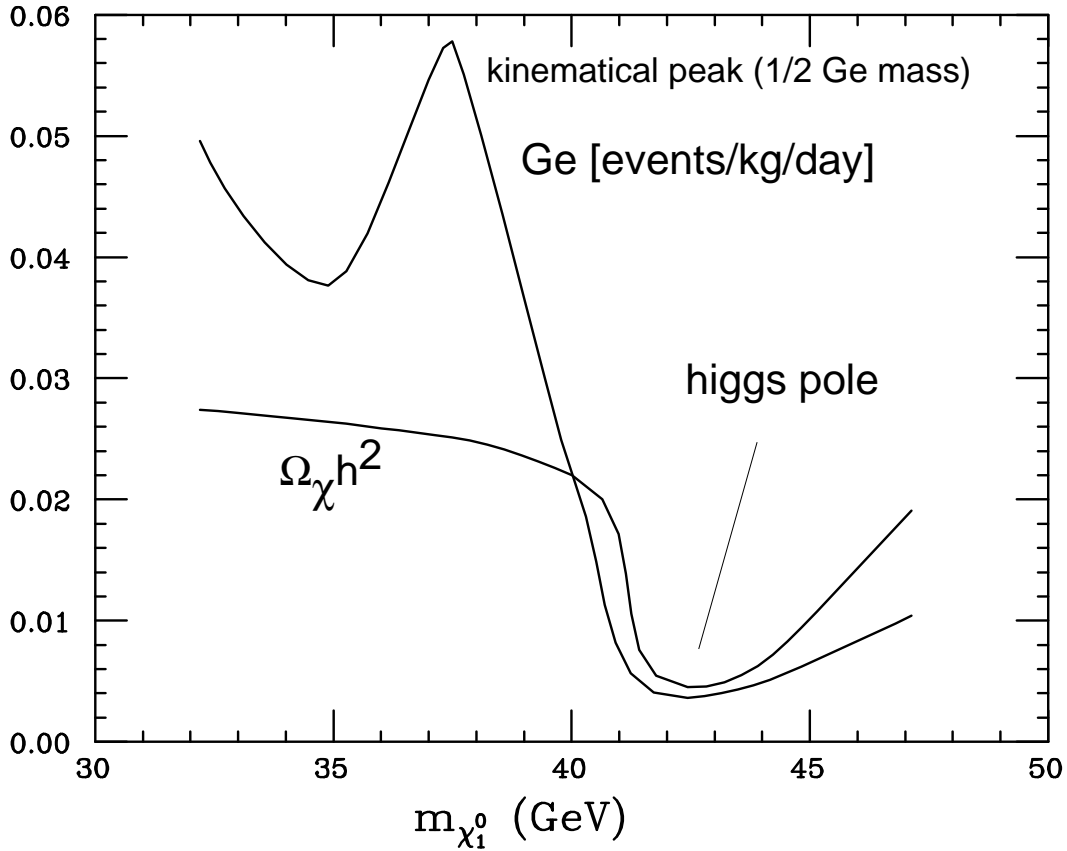}
\vspace{-2cm}
\caption{The relic density of neutralinos as a function of the LSP mass. Also
shown is the predicted detection rate in the Ge detector in events/kg/day.}
\label{fig:lsp}
\end{figure}

In our model the lightest supersymmetric particle (LSP) is stable (as R-parity
is unbroken) and will contribute to the cold dark matter in the Universe. The
calculation of the relic abundance of neutralinos yields \cite{zero}
\begin{equation}
\Omega_\chi h^2\lsim0.025\ ,
\end{equation}
where $h$ is the scaled Hubble parameter. Such magnitude of neutralino relic
density is of interest in models with a sizeable cosmological constant
\cite{CC}, where one could expect
\begin{equation}
\Omega_\nu(0.3)+\Omega_\chi(0.1)+\Omega_\Lambda(0.6)=1\ ,
\end{equation}
with a hot dark matter component (neutrinos, see Sec.~\ref{sec:neutrinos}), a
cold dark matter component (neutralinos), and a cosmological constant
component. Our neutralinos would populate the galactic halo and could be
detected via scattering off nuclei \cite{lspd}.
In present-day cryogenic Germanium detectors, the present sentivity is of 0.1
events/kg/day (eventually expected to improve to 0.01 events/kg/day). In
Fig.~\ref{fig:lsp} we show the relic density of neutralinos as a function of
the LSP mass, showing a dip when the higgs-boson pole is encountered. We also
show the predicted detection rate in the Ge detector, with its characteristic
kinematical peak when the LSP mass is half of the Ge mass.

\section{Conclusions}
We have reviewed the features that make flipped SU(5) and very nice
supersymmetric unified theory, including recent developments that allow a
natural reduction in the prediction for $\alpha_s$, which is consistent with
the whole range of presently experimentally allowed values. The real power of
flipped SU(5) lies, however, in its being string derivable, as then many more
predictions follow. This string model suppresses naturally the tree-level and
one-loop contributions to the vacuum energy, and has a distinct prediction
for the supersymmetry-breaking parameters. The anomalous $\rm U_A(1)$ plays
a crucial role in the one-loop suppression, and is conjectured to suppress
the vacuum energy to all orders in string perturbation theory. Another new
development consists of a new scenario for flipped string unification, wherein
the ``LEP" scale (where the SU(3) and SU(2) gauge couplings unify) is obtained
by means of having the SU(5)$\times$U(1) breaking vevs participate in the
anomalous $\rm U_A(1)$ cancellation mechanism. The SU(5) and U(1) gauge
couplings then unify at the string scale. This scenario predicts the existence
of a new (\r{10},\rb{10}) complete pair at the scale $10^{8-9}\GeV$, that can
be obtained from hidden sector matter condensation. Among the tests
of flipped SU(5) we have its prediction for $\alpha_s$ and the proton lifetime
into the $e^+\pi^0$ mode, its prediction for the top-quark mass, its prediction
for the superparticle masses and the rates at which they could be produced in
direct and indirect processes, and its prediction for the hot ($\nu_\tau$) and
cold ($\chi^0_1$) dark matter in the Universe and its detection.

\section*{Acknowledgments}
The work of J.~L. has been supported in part by DOE grant DE-FG05-93-ER-40717.
The work of D.V.N. has been supported in part by DOE grant DE-FG05-91-ER-40633.

\end{document}